# Half-metallic Ferromagnets, Spin Gapless Semiconductors, and Topological Semimetals Based on Heusler Alloys

**V.V. Marchenkov[a,b],\* and V.Yu. Irkhin[a]**

[a] *M.N. Mikheev Institute of Metal Physics, Ural Branch, Russian Academy of Sciences, Ekaterinburg, 620108 Russia*

[b] *Ural Federal University, Ekaterinburg, 620002 Russia*

\**e-mail: march@imp.uran.ru*

A review of theoretical and experimental studies of the electronic structure, electronic and magnetic properties of various systems of Heusler alloys in the states of a half-metallic ferromagnet, a spin gapless semiconductor, and a topological semimetal is presented. These substances have unusual, highly sensitive to external influences, magnetic and electronic characteristics, which is associated with the presence of energy gaps and exotic excitations in them. The features of the behavior and evolution of the electronic structure and properties in each of these states, as well as during the transition between them, are considered. The possibility to purposefully control the properties of such materials is prospective for their practical application.

## CONTENTS

## 1. Introduction

The search, creation, and comprehensive theoretical and experimental study of novel magnetic materials and nanostructures on their basis are of great interest and relevant for practice, since these systems can be used in modern micro- and nanoelectronics, as well as spintronics. Heusler alloys in the half-metallic ferromagnet, spin gapless semiconductor, and topological semimetal states belong to such materials: they can realize a high degree of spin polarization of current carriers and, consequently, create a spin current.

In the modern scientific literature, there are several reviews on half-metallic ferromagnets (HMFs), spin gapless semiconductors (SGS), and topological semimetals (TSMs), which, in particular, are realized in Heusler alloys (see, e.g., [1-3]). However, these reviews were either published more than 10 years ago or are devoted to describing a wide range of functional characteristics of Heusler alloys without paying sufficient attention to the HMF, SGS, and TSM states. Therefore, the purpose of this review is to discuss the current state of the art in the field of theoretical and experimental studies of Heusler alloys in these states.

The study of half-metallic ferromagnets as a class of magnetic materials began with a theoretical prediction in 1983 by R. de Groot, who performed band calculations of NiMnSb [4]. The main feature of HMFs is the presence of a gap at the Fermi level for electronic states with one of the spin projections. In the simple mean-field approximation, this means 100% spin polarization of the current carriers; however, taking into account correlation effects and incoherent states significantly complicates the physical picture. Thus, experimental confirmation of the HMF state is not an easy task. Studying kinetic properties can help here. In particular, in the scattering mechanisms, two-magnon processes are dominating, which leads to anomalous temperature dependences of the kinetic properties.

From the standpoint of practical applications, it is important to study systems with properties close to the classical well-studied semiconductors, such as EuO, EuS, $HgCr_2Se_4$, etc. Degenerate ferromagnetic semiconductors are also HMF systems. Recently, a certain change in terminology has taken place: now they talk about half-metallic ferromagnetism in high-quality single crystals of doped $HgCr_2Se_4$ (see, e.g., [5]).

A class of materials close to HMFs are the SGS materials, predicted in 2008 [6]. They also have a wide ($\Delta E \sim 1$ eV) energy gap near the Fermi level for one projection of the spin of current carriers, and, for carriers with opposite spin directions, there is a zero gap, similar to that observed in classical gapless semiconductors [7].

Recently, exotic materials with nontrivial topology – topological semimetals – have been discovered, in which new quantum phenomena arise, in particular, those associated with Dirac-type "massless" fermions. The TSMs were experimentally studied mainly in the case of transition metal chalcogenides $MX_2$ (M = Mo, W, V, etc.; X = Te, S, Se, etc.) [8], but, by now, there appeared the first publications on the observation of TSM states also in Heusler alloys [9].

Since many Heusler alloys with the general formula $X_2YZ$ (X and Y are usually 3d elements and Z are s and p elements of the Periodic table) belong to HMF, SGS, and TSM materials [3, 6, 9 – 11], the study of the electronic structure and magnetic state of such alloys is very promising. In this case, the position and width of the band gap can be quite different in various systems. These parameters can be varied by changing the 3d, s, and p elements in Heusler alloys $X_2YZ$, thereby changing the electronic properties.

Usually, the rise of HMF, SGS, and TSM states is judged from the results of calculations of the electronic band structure and/or the data of optical measurements. However, they must also manifest themselves in the behavior of electron transport properties (electro- and magnetoresistivity, Hall effect, thermo emf, etc.) and magnetic characteristics. Therefore, along with the first-principles calculations, we analyze the correlations between all these properties.

In this review, we consider specific Heusler alloys. Experimental and theoretical studies in this direction enable one to describe the evolution of the electronic structure and properties of Heusler compounds from a unified standpoint, to understand the features of the manifestation of the states of a half-metallic ferromagnet, spin gapless semiconductor, and topological semimetal, their generality and differences.

After a general brief introduction to the physics and crystallography of Heusler alloys (Section 2), we consider the classes of HMF, SGS, and TSM compounds (Sections 3 – 6). In Section 7, we conduct a general discussion of the possible realization of such states and transitions between them in Heusler alloys.

## 2. Heusler alloys. Types of crystal structure

In 1903, German chemist Fritz Heusler discovered the Heusler alloys [12], in particular, $Cu_2MnAl$, which exhibits strong ferromagnetic properties with a high Curie point (although each of its constituent elements – Cu, Mn and Al – is not a ferromagnet). Since then, about 1500 different Heusler alloys with diverse functional properties have been found (Fig. 1). Among them, there are compounds with the shape memory effect [13 – 15] and the giant magnetocaloric effect [15 – 17], thermoelectrics [18 – 20], alloys with unusual thermal [20 – 22] and semiconducting properties [19,20, 23], superconductors [3, 24, 25], and many others.

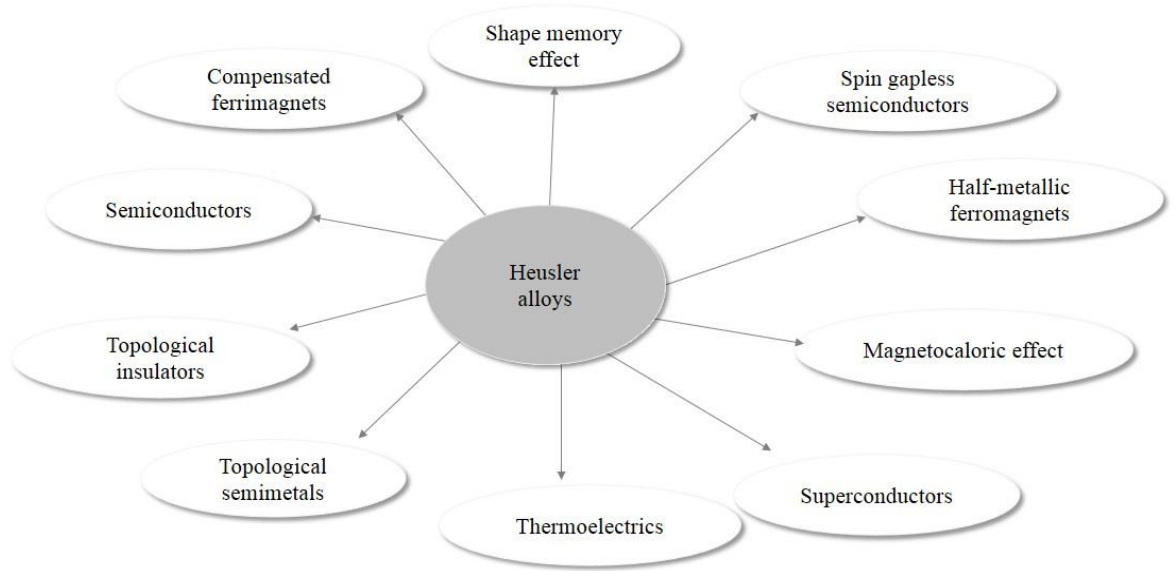

**Fig. 1.** Various types of Heusler alloys and their functional properties.

Heusler alloys are intermetallic compounds with the chemical formula *XYZ* (half Heusler alloys), $X_2YZ$ (full Heusler alloys) and $XX'YZ$ (quaternary Heusler alloys). These compounds are formed from many combinations of elements, where *X*, *X'*, and *Y* are usually transition, rare-earth, or noble metals (e.g., Ni, Co, Pd, Cu, Pt, Au, Mn, Fe, Co, Ti, V, Zr, Nb, Hf, etc.) and *Z* are elements of the IIIB – VB groups: Al, Ga, In, Si, Sn, Ge, Sb, etc.

Figure 2 schematically shows the crystal cells of the full, half, and inverse Heusler alloys with the structures $L2_1$, $C1_b$, and $X_A$, respectively.

Heusler ternary intermetallic compounds with the general formula $X_2YZ$ during crystallization are first ordered into a high-temperature austenite phase with an fcc lattice, consisting of eight unit cells of the *B2* (CsCl) type. Depending on the elemental composition that forms the basic formula of the compound and the coordinates of specific atoms, three types of ordered superstructures are distinguished. Most of the $X_2YZ$ compounds belong to the so-called full Heusler alloys $L2_1$ (of the $Cu_2MnAl$ type) with a space group $Fm3m$ (Fig. 2a).

Another group of compounds, called half-Heusler alloys, with the formula *XYZ* has a $C1_b$-type (MgAgAs-type) (Fig. 2b) structure. In the $L2_1$ structure, all four sublattices in the fcc lattice are occupied by *X*, *Y*, and *Z* atoms, while, in the $C1_b$-type structure, some of the positions are vacant.

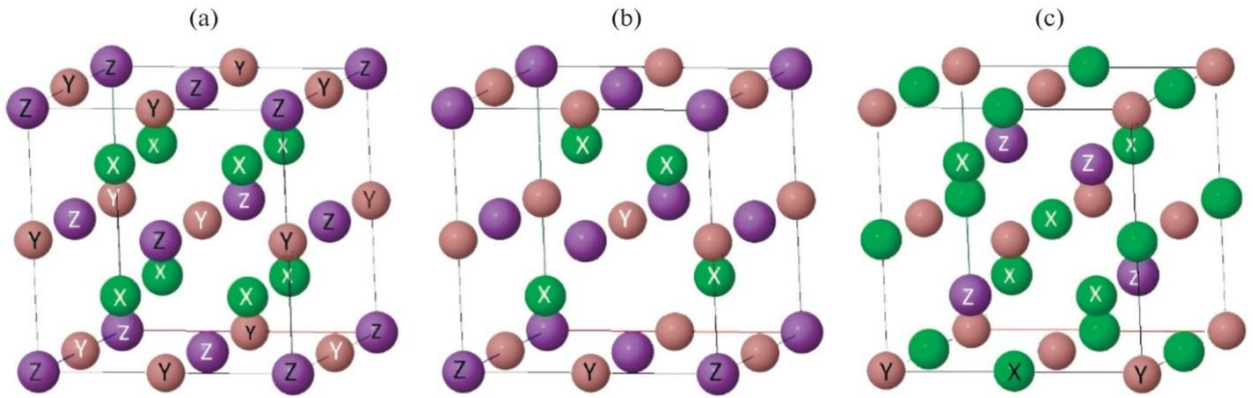

**Fig. 2.** Crystal cells of Heusler compounds: (a) full alloy $X_2YZ$ (structure $L2_1$), (b) half alloy *XYZ* (structure $C1_b$), and (c) inverse alloy $X_2YZ$ (structure $X_A$)

The third family is formed by the $Hg_2CuTi$-type compounds, called the $X_A$ inverse Heusler structure. There is a similarity with the main Heusler structural type $L2_1$ with the formula $X_2YZ$, but, in the inverse version, the atomic number of the element *Y* is higher than that of the element *X* (Fig. 2c).

The formation of the structure of a Heusler alloy upon solidification of the melt is possible either through a completely disordered phase *A*2 ($A2 \rightarrow L2_1$) or through an intermediate partially ordered phase *B*2. Upon further cooling or under the action of a load and a magnetic field, the alloys can undergo martensitic transformations with the formation of low-symmetry phases, both modulated (multilayer) and unmodulated; for example, the tetragonal bct structure $L1_0$ (*I4/mmm*) or orthorhombic or monoclinic multilayer structures 10*M*, 14*M*, 6*O*,

*3R*, etc. Below, Fig. 3 shows tetragonal distortions of the austenite lattice for full and inverse Heusler alloys.

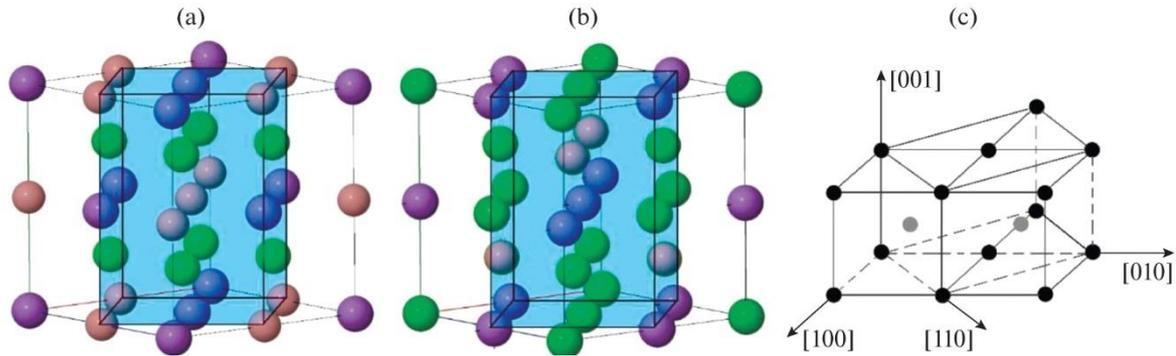

**Fig. 3.** Crystal structure of full Heusler alloy $X_2YZ$ with a $Cu_2MnAl$-type lattice (a) and an inverse Heusler alloy with a $Hg_2TiCu$-type lattice (b). Tetragonal cells are highlighted in blue. (c) Schematic representation of the structural mechanism of rearrangement of a cubic lattice into a tetragonal one according to Bain.

It should be noted that the structural transitions from austenite to martensite, observed in ferromagnets, as well as in other nonmagnetic metastable austenitic alloys, are also preceded by various pretransition premartensitic phenomena: softening of the elastic moduli; the formation of soft phonon modes in the lattice, anomalous in the amplitude of atomic displacements and visualized from the data of neutron diffraction; X-ray phase analysis; observation of tweed contrast and diffuse scattering in images obtained by transmission electron microscopy, etc.

## 3. Simple models of ferromagnets, half-metallic ferromagnets, spin gapless semiconductors, and topological semimetals

Let us consider simple qualitative models of conventional ferromagnets, half-metallic ferromagnets, spin gapless semiconductors, and topological semimetals. Figure 4 schematically shows the corresponding densities of states.

In a ferromagnetic metal at $T = 0$, all electronic states below the Fermi level $E_F$ are occupied and, above the Fermi level, vacant and magnetic ordering leads to a slight polarization of the current carriers (Fig. 4a). The cases of half-metallic ferromagnets and spin gapless semiconductors have the following features. The HMFs are characterized by the presence of a gap at the Fermi level for spin down electronic states, which is absent for spin up carriers (Fig. 4b). In a simple picture, this means 100% spin polarization of the current carriers; however, in real systems, the situation is much more complicated, and it is necessary to take into account correlation effects and incoherent states. For SGSs (Fig. 4c), as in the case of HMFs, there is a finite gap for one of the spin projections; for the other projection, the gap is zero, similar to that observed for classical gapless semiconductors [7].

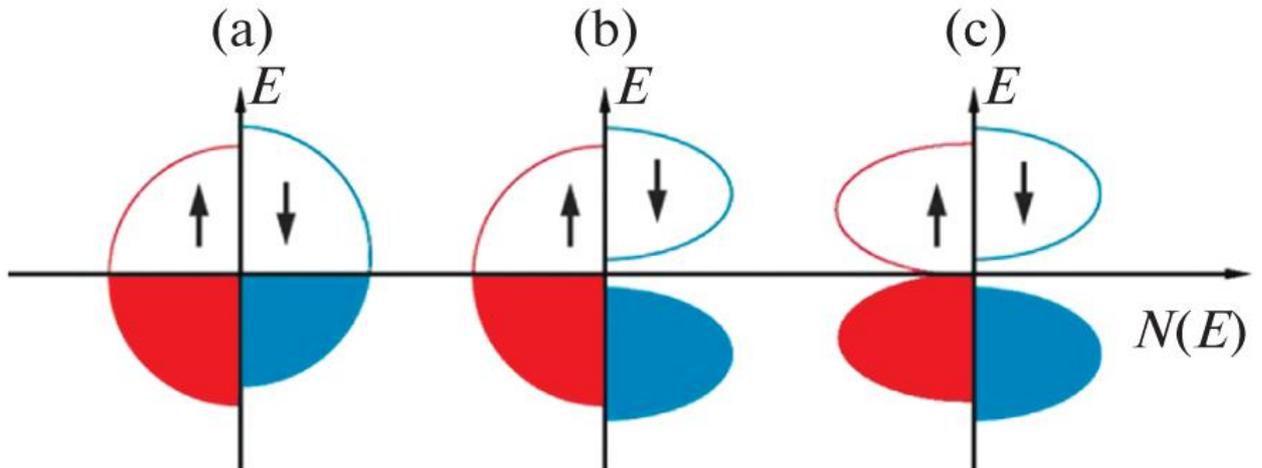

**Fig. 4.** Schematic representation of the density of states of (a) a ferromagnet, (b) a half-metallic ferromagnet, and (c) a spin gapless semiconductor. The arrows indicate the directions of the spins for the electronic states. In contrast to case (b), in case (c), the density of states at the Fermi level is zero for both spin projections.

In topological materials (Fig. 5), as a result of strong spin-orbit interaction, leading to inversion of the conduction and valence bands, a nontrivial topology of the electronic band structure arises. Such materials include topological insulators and Dirac and Weyl semimetals. In the volume of topological insulators, there is a characteristic energy gap (Fig. 5a) and there are "metallic" states on the surface. In the volume of the topological Dirac and Weyl semimetals, there is also a gap, which appears due to the strong spin-orbit coupling, except for the points of intersection of the bands at the Dirac (Fig. 5b) and Weyl (Fig. 5c) points, respectively.

Near these points, the band dispersion in all three directions of momentum space is linear and low-energy excitations can be described by the Dirac or Weyl Hamiltonians.

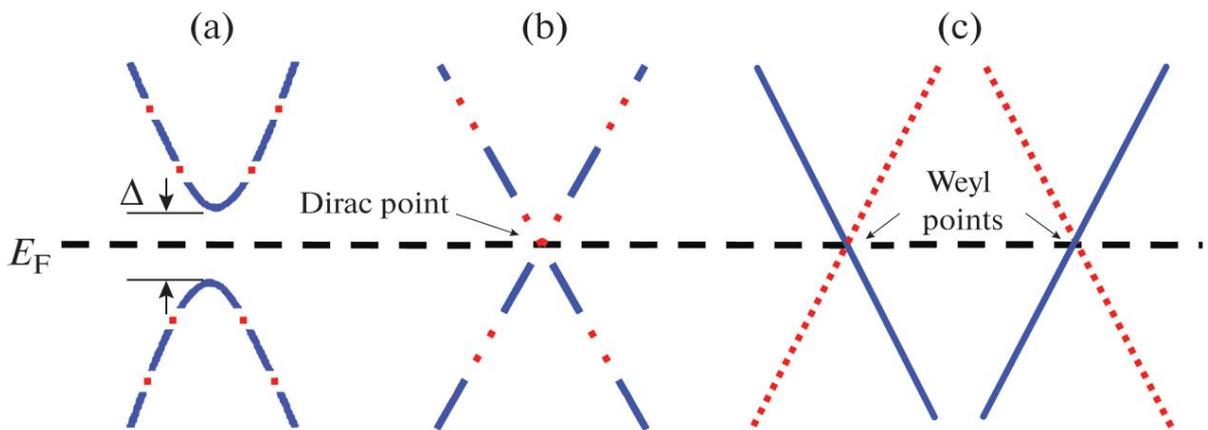

**Fig. 5.** Schematic representation of the band structure of (a) massive Dirac fermions with a gap $\Delta$ and (b) massless Dirac and (c) Weyl fermions. The latter arise during the decay of the Dirac point. Curves and lines consisting of solid and dashed lines represent doubly degenerate zones, and only those of solid or dashed ones represent nondegenerate zones.

In Sections 4-6, these classes of materials will be considered in more detail.

## 4. Half-metallic ferromagnets

The case of "strong" ferromagnetism with a large spin splitting is the opposite of weak itinerant ferro-magnets. In the Stoner theory, a ferromagnetic solution was considered in which the spin splitting is large, so that one spin subband is empty and the other is partially filled (Wohlfarth's solution; see, e.g., [1, 2]). It was believed that this situation could correspond to pure ferromagnetic metals of the iron group, but band calculations refuted this assumption.

At the same time, band calculations led to the discovery of real magnets similar to strong Stoner ferro-magnets. Calculations by de Groot et al. of the band structures of Heusler alloys NiMnSb [4] and PtMnSb [4, 27, 28] with the *C1_b* crystal structure demonstrate that the Fermi level for one of the spin projections is in the energy gap. Since these systems for one of the spin projections behave like insulators, they were called "half-metallic ferromagnets". Later, a similar picture was obtained for CoMnSb [29] and ferrimagnetic FeMnSb [30]. For CrMnSb, a state was found called "half-metallic antiferromagnet" [27]. However, it should be rather considered a compensated ferrimagnet; in the case of Heusler alloys, such states will be discussed in Section 7.

Interest in HMFs was primarily due to their unique magnetooptical properties [27, 28], which are closely related to the electronic structure near the Fermi level (the absence of states with one spin projection), that leads to a strong asymmetry of optical transitions. In particular, the magnetooptical Kerr effect was observed in the PtMnSb alloy, for which this effect is especially large due to strong relativistic interactions in the Pt atom.

Until now, the presence or absence of the HMF state is usually judged by the results of band calculations of the density of electronic states (see, e.g., [31 – 33] and references therein), when there is a finite density of states for spin up carriers and a gap in $E_F$ for spin down carriers.

The magnitude of the spin polarization $P$ can be experimentally determined using one of the direct methods: angle-resolved X-ray photoemission spectroscopy (ARPES). In [34, 35], this method was used to experimentally study thin films of the Heusler alloy $Co_2MnSi$. It was shown that $P$ reaches a value of about 93% even at room temperatures [35].

*4.1. Mean-field approximation and correlation effects. taking into account non-quasiparticle states*

In addition to qualitative consideration and calculations of the band structure, for the systems under discussion, theoretical approaches have been developed using models with strong electron-electron correlations [1, 2]. The main ones are the one-band Hubbard model, which takes into account the intrasite Coulomb interaction, and the two-band *s – d* exchange model, which considers the interaction of localized moments with mobile conduction electrons. The Hamiltonian of the latter has the form

$$H = \sum_{k\sigma} t_k c^\dagger_{k\sigma} c_{k\sigma} + \sum_q J_q S_{-q} S_q - I \sum_{i\sigma\sigma'} (S_i \sigma_{\sigma\sigma'}) c^\dagger_{i\sigma} c_{i\sigma'} \quad (1)$$

where $t_k$ is the band energy, $c_{i\sigma}$ and $\mathbf{S}_i$ are the operators of conduction electrons and localized spins, $\sigma$ are Pauli matrices, $\mathbf{J}_q$ are the Fourier transforms of exchange integrals between localized $d$ states, and $I$ is the parameter of the $s - d(f)$ exchange interaction.

Due to the special band structure of HMFs, an important role in them is played by non-quasiparticle (NQP) states arising near the Fermi level due to the electron correlation effects [2, 36, 37]. These states go beyond the framework of the mean-field theory (in particular, the Stoner theory), in which there are only two quasiparticle spin-split spin subbands.

The appearance of NQP states in the band gap near the Fermi level is one of the most interesting correlation effects characteristic of HMFs. The origin of these states is associated with spin-polaron processes: low-energy electron spin down excitations, forbidden for HMFs in the standard one-particle scheme, are possible as a superposition of excitations of spin up electrons and virtual magnons at $T = 0$ and real magnons at finite $T$ (Fig. 6). They are formally described by the imaginary part of the electron self-energy, which is determined by the convolution of the magnon Green's function and the Green's function of carriers with the opposite spin projection. Thus, in the energy gap, states with both spin projections can arise.

At zero temperature, the density of these non-quasiparticle states vanishes at the Fermi level, but sharply increases on an energy scale on the order of the characteristic magnon frequency $\tilde{\omega}$. However, with an increase in temperature, this contribution to the density of states increases proportionally to the deviation of the mean spin projection $\langle S_z \rangle$ from the maximum value of $S$, i.e., according to Bloch's $T^{3/2}$ law (not exponentially, as it would be in the mean-field theory). This is important for the behavior of the spin polarization of conduction electrons.

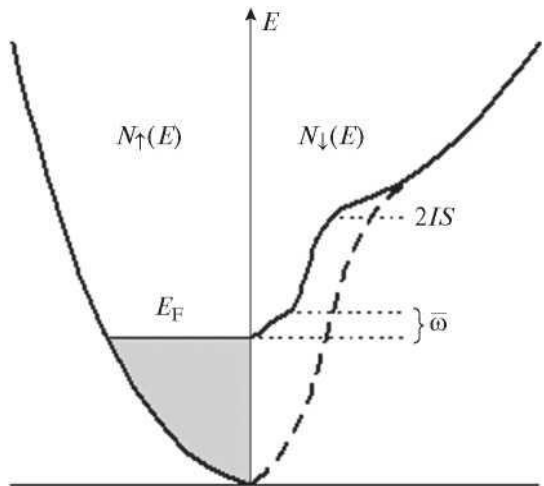

**Fig. 6.** Density of states in the *s-d* model of HMF with the exchange parameter $I > 0$, $2IS$ is the spin splitting, and ra is the characteristic magnon frequency. (Dashed line) The "tail" of non-quasiparticle states, proportional to $T^{3/2}$.

Non-quasiparticle states were first considered theoretically by Edwards and Hertz [38]

within the broad-band Hubbard model for itinerant electron ferromagnets. It was shown later [39] that, for the narrow-band Hubbard model with infinite Coulomb repulsion, the entire spectral weight for one spin projection belongs to non-quasiparticle states, which is of decisive importance for the problem of stability of strong ferromagnetism and for an adequate description of the corresponding excitation spectrum.

Non-quasiparticle states in the *s-d*-exchange model of magnetic semiconductors were considered in [40]. It was shown that, depending on the sign of the *s-d* exchange integral *I*, these states can form either only below the Fermi energy $E_F$ or only above it. Later it turned out that half-metallic ferromagnets are natural systems for theoretical and experimental studies of such states. As an example of very unusual properties of NQP states, we note that they can contribute to the term linear in temperature in the electronic heat capacity [41, 42], despite the fact that their density at $E_F$ vanishes at $T = 0$.

The existence of NQP states on the HMF surface was predicted in [43]. They can be detected by surface-sensitive methods such as ARPES [44] or spin-polarized scanning tunneling microscopy [45]. Such states may be important when considering topological motives, since surface states play a fundamental role in topological materials.

The density of non-quasiparticle states was calculated from first principles for NiMnSb [46]. Recently, the HMF behavior and the presence of NQP states were directly detected in $CrO_2$ using volume-sensitive spin-resolved photoemission spectroscopy [47]. This work also revealed a significant temperature dependence of the mean spin polarization (about 100% at 40 K and 40% at 300 K). At 100 K, a special depolarization at the Fermi level was discovered, which can be associated with NQP states.

Non-quasiparticle states make a significant contribution to the magnetic and transport properties [36]. A theoretical study of spin transport in HMFs at finite temperatures taking into account the band minorirty NQP spin states was considered in [48], where a spin Hall conductivity proportional to $T^{3/2}$ was found.

*4.2. Kinetic properties*

In a simple consideration, a HMF conductor can be represented as a system of two parallel-connected conductors [49, 50]. One of them is a conduction channel for spin up carriers, and the other, for spin down carriers. At temperatures much below the gap, only the first of these conductive channels is active.

The spin polarization *P* at the Fermi level $E_F$ of the material is defined as

$$P = (N_\uparrow(E_F) - N_\downarrow(E_F))/(N_\uparrow(E_F) + N_\downarrow(E_F)), \qquad (2)$$

where $N_\uparrow(E_F)$ and $N_\downarrow(E_F)$ are the densities of states of spin up and spin down electrons at the Fermi level $E_F$, respectively. In ferromagnets or ferrimagnets, *P* has a finite value below the Curie temperature $T_C$. In the mean-field theory for HMFs, up to exponentially small contributions, either $N_\uparrow(E_F) = 0$ or $N_\downarrow(E_F) = 0$, so that, below $T_C$, the electrons at the Fermi level are almost completely polarized (*P* = 100%). However, with allowance for the effects of electron-magnon interaction, the spin polarization exhibits complex temperature behavior (roughly speaking, as that of the relative magnetization) due to the contribution of the non-quasiparticle states [2, 36, 37] (see also Fig. 6).

At room temperature, a high degree of spin polarization of current carriers can be realized in HMF alloys with high Curie temperatures $T_C$. Such materials include cobalt-based Heusler compounds $Co_2YZ$, where large values of $P$ are observed [51, 52]. In particular, a high degree of spin polarization at room temperature was found in the alloys $Co_2MnSi$, $Co_2FeSi$, and $Co_2Cr_{0.6}Fe_{0.4}Al$ [35, 53-55].

In ordinary ferromagnetic metals, the magnetic contribution to the resistivity is determined by the one-magnon processes:

$$\rho(T) \sim T^2 N_\uparrow(E_F) N_\downarrow(E_F) \exp(-T/T^*), \qquad (3)$$

where $T^* \sim q_1^2 T_C$ is the characteristic scale for these processes, $q_1 \sim \Delta/v_F$, and $\Delta = 2IS$ is the spin splitting (energy gap). In half-metallic ferromagnets, this contribution is absent, since $N_\downarrow(E_F) = 0$. Two-magnon scattering processes [56] lead to a power-law temperature dependence of the resistivity, $\rho(T) \sim T^n$, as well as to a negative linear magnetoresistivity. In this case, $n = 9/2$ for $T > T^{**}$ and $n = 7/2$ for $T > T^{**}$, where $T^{**} \sim q_2^2 T_C$. In a simple one-band HMF model, where $E_F < \Delta$, we have $q_2 \sim (\Delta/W)^{1/2}$ ($W$ is the band width). Generally speaking, $q_2$ can be sufficiently small provided that the band gap is much smaller than $W$, which is typical of real HMF systems [2, 56].

An important feature of half-metallic ferromagnetism is an unusual behavior of the nuclear spin-lattice relaxation rate $1/T_1$. The point is that due to the absence of the density of states at the Fermi level for one of the spin projections, there is no usual Korringa contribution, linear in temperature and proportional to $TN_\uparrow(E_F)N_\downarrow(E_F)$. Thus, the dominant contribution must be the contribution of two-magnon processes, the calculation of which gives $1/T_1 \sim T^{5/2}$ [1].

The experimental observation of the dependence $\rho(T) \sim T^{9/2}$, predicted in [56], was reported in [57], where the temperature dependences of the electrical resistivity of Heusler alloys based on $Co_2FeSi$ were studied with the replacement of half of the Si atoms by Al, Ga, Ge, In, and Sn. The authors of [57] showed that, at low temperatures up to 50 – 80 K, the dependence $\rho(T)$ of the studied alloys can be described by the formula $\rho(T) = \rho_0 + aT^2 + bT^{9/2}$, where $\rho_0$ is the residual resistivity and $a$ and $b$ are coefficients.

In [58], where the temperature dependences of the resistivity $\rho(T)$ of a $Co_2FeSi$ single crystal were measured in magnetic fields from 0 to 150 kOe, it was shown that there are three temperature ranges in which the resistivity depends differently on temperature and magnetic field (Fig. 7): (1) below 30 K, $\rho(T) \propto AT^n$ with $n \approx 2$ and coefficient $A \propto H^2$; (2) in the range from 30 to 60 K, $\rho(T) \propto CT^n$ with $n \approx 4$ and the coefficient $C \propto -^lH$; 3) above 65 K, $\rho(T) \propto BT^n$ with $n \approx 2$ and the coefficient $B \propto H^{-2}$

The experimental results obtained (Fig. 7) show that, in the temperature range 30 K < $T$ < 60 K, a power-law temperature dependence of the resistivity with a large exponent and linear negative magnetoresistivity is observed. Apparently, this is a manifestation of two-magnon scattering processes: they are the main mechanism of carrier scattering, which determines the behavior of the electro- and magnetoresistivity of the alloy in this temperature range.

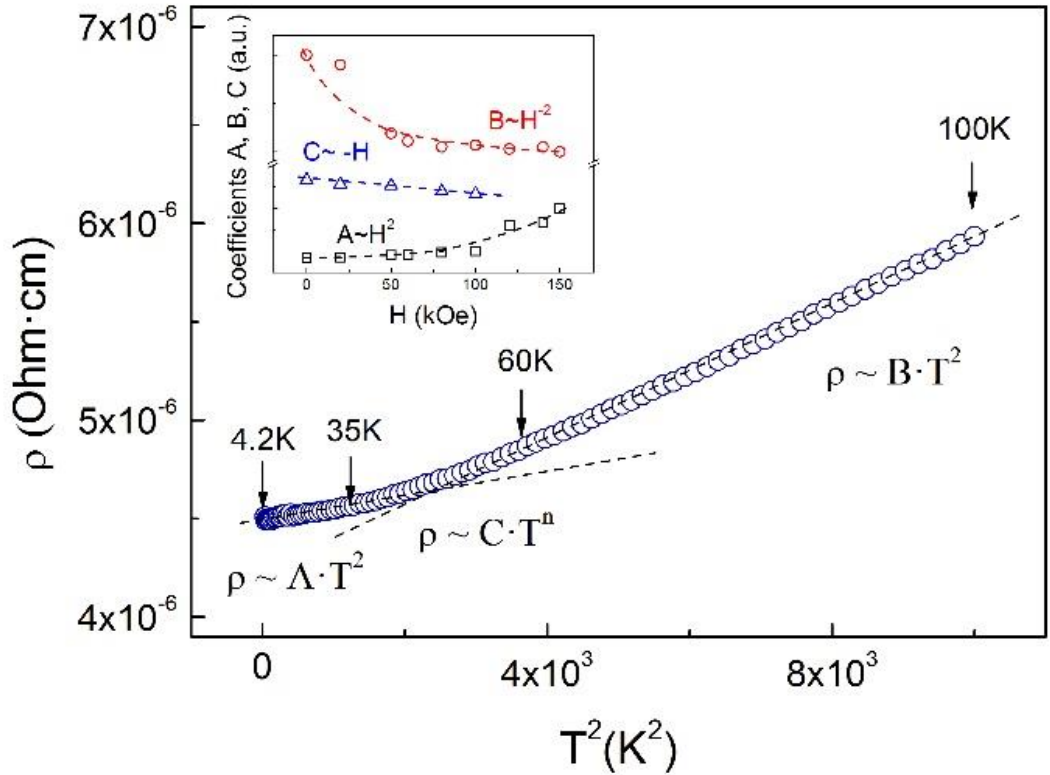

**Fig. 7.** Electrical resistivity of the $Co_2FeSi$ alloy vs. the square of temperature. (Inset) Field dependences of the coefficients *A, B,* and *C*.

Negative magnetoresistivity with a linear dependence on the field was observed in many HMF systems (see, e.g., [59 – 63]). Figure 8 shows the field dependences of the magnetoresistivity $\Delta\rho/\rho_0$ of the $X_2YZ$ alloys, obtained using the data from [59 – 63]. It can be seen that, in strong magnetic fields, $\Delta\rho/\rho_0$ is negative and changes according to a law close to linear in the field. This behavior may be a consequence of two-magnon scattering processes [56], which are especially well manifested in HMFs at temperatures much below the gap.

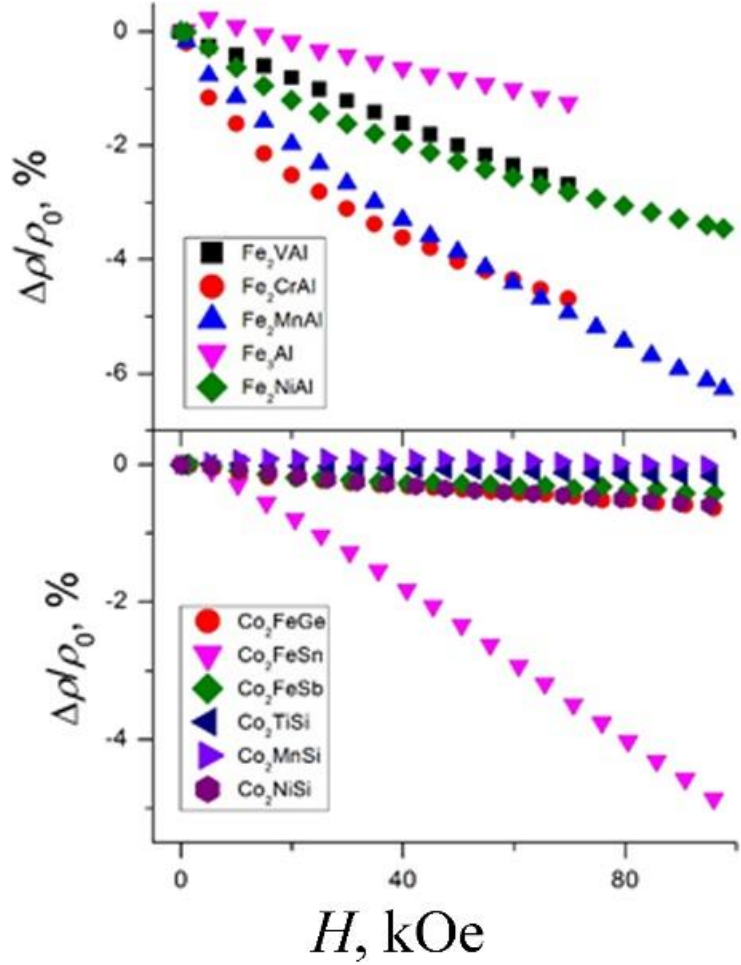

**Fig. 8.** Field dependences of magnetoresistivity $\Delta\rho/\rho$ of alloys Fe$_2$YAl, Co$_2$FeZ, and Co$_2$YSi at $T$ = 4.2 K.

The authors of [53], where they measured the electrical resistivity and studied the galvanomagnetic properties of the Co$_2$FeSi single crystal, presented the temperature-dependent part of the resistivity as the sum of contributions from the electron-phonon, $\rho_{ph}$, and electron-magnon, $\rho_M$, scattering processes. The latter was written in the form

$$\rho_M(T) = cT^2 \cdot e^{-\Delta/T} \quad (4)$$

where $c$ is the parameter of the electron-magnon scattering efficiency and $\Delta$ is the gap width in energy units. By fitting the parameters, the authors described the experimental curve and determined the gap width $\Delta$ = 103 K, which corresponds to an energy $k_B\Delta$ = 8.9 meV. According to the authors, the gap width is much smaller in comparison with theoretical calculations (its predicted width is greater by more than an order of magnitude). Therefore, a more reasonable interpretation should take into account two-magnon scattering processes, which give not exponentially small, but power-law contributions to the resistivity.

*4.3. Calculation of the electronic band structure*

The studies of the magnetic properties of Heusler alloys $Co_2YZ$ showed that the Slater-Pauling rule [51], which describes the relationship between the number of valence electrons and the magnetic moment, is well satisfied in them (Fig. 9). Theoretically, it allows one to predict the total magnetic moment of such alloys.

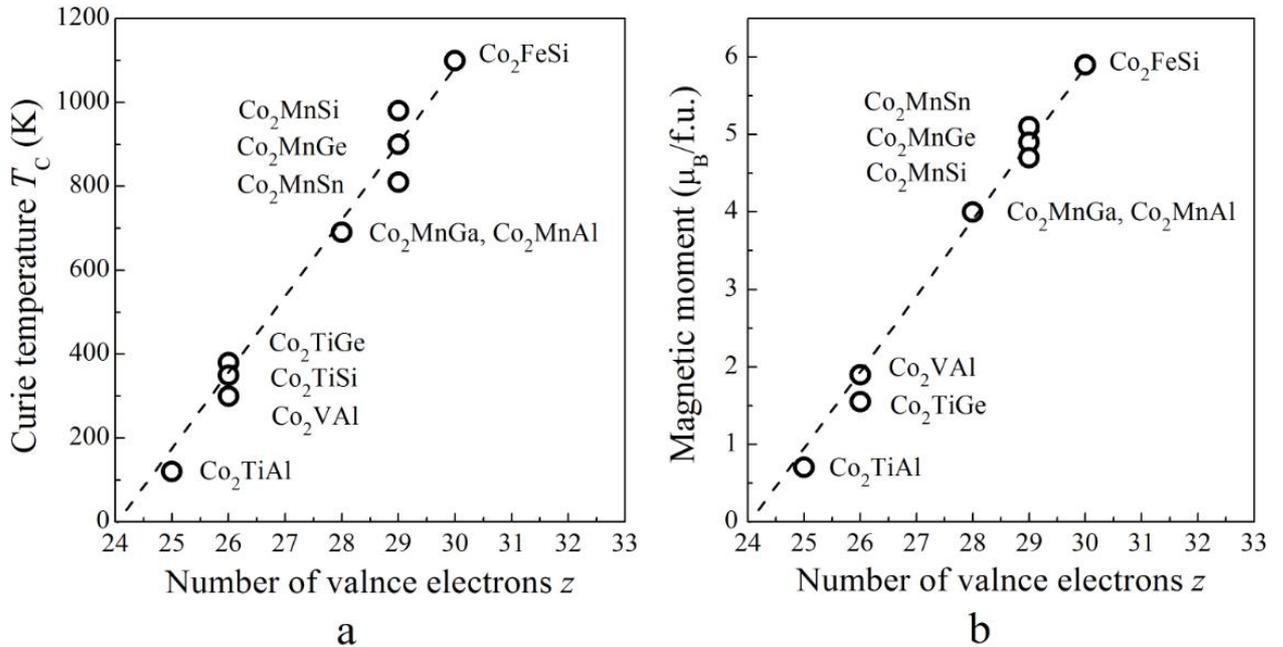

**Fig. 9.** (a) Curie temperature and (b) total magnetic moment of some cobalt-based Heusler alloys vs. the total number of valence electrons.

In half-metallic ferromagnets, the Slater-Pauling rule can be written as follows [51]:

$$M = z - 24, \qquad (5)$$

where $M$ is the magnetic moment per formula unit in Bohr magnetons and $z$ is the total number of valence electrons.

For the $Co_2YZ$ alloys, an almost linear dependence of the Curie temperature $T_C$ on the number of valence electrons is observed (Fig. 9b). The maximum value of $T_C$ takes place in the case of the largest number of valence electrons. As can be seen from Fig. 9 (see also [53]), the $Co_2FeSi$ alloy has a maximum Curie temperature $T_C = 1100$ K and a maximum magnetic moment of 5.97 $\mu_B$/f.u. at 5 K.

The situation of qualitatively different states for spin up and down that is realized in HMFs is nontrivial for the general theory of itinerant magnetism [1]. The scheme of the formation of a half-metallic state in $XMnZ$ and $X_2MnZ$ Heusler alloys with the structures $C1_b$ and $L2_1$ can be described as follows [4, 28, 64, 65]. Neglecting the hybridization of the atomic states $X$ and $Z$, the $d$ band of manganese is characterized by a wide energy gap between the bonding and antibonding states. Due to the strong intra-atomic (Hund's) exchange for manganese ions in the ferromagnetic state, the subbands with spins up and down are significantly spaced apart.

One of the spin subbands is close to the *p* band of the ligand and, therefore, the corresponding gap is partially or completely blurred by $p - d$ hybridization. The energy gap in the other subband is preserved and, under certain conditions, can coincide with the Fermi level, which gives the HMF state.

For the $C1_b$ structure, we have a true gap, and, for the $L2_1$ structure, a deep pseudogap. This is associated with a significant change in the nature of $p - d$ hybridization (especially between the states of the *p* and $t_{2g}$ nature) in the absence of an inversion center, which is the case for the $C1_b$ structure. Thus, the latter structure is more favorable for the HMF state. According to [66], similar factors are responsible for the gap in the partial density of states for one of the manganese positions (Mn(I)) in the $Mn_4N$, the structure of which is obtained from the structure of $X_2MnZ$ by removing some atoms. A qualitatively similar mechanism, which is based on strong Hund exchange and hybridization between the *d* states of chromium and *p* states of oxygen, is considered in [67] for $CrO_2$.

The stability of the ferromagnetic state is a consequence of the difference in $p - d$ hybridization for states with opposite spin projections (see discussion in [64]). To describe this situation, Kuebler et al. [64] introduced the term "covalent magnetism" and emphasized the difference between the spectral picture and the Stoner model, where the densities of states with opposite spin projections differ in a shift by a constant spin splitting.

In [43, 69 – 76], calculations of the electronic structure of various systems of ternary and quaternary Heusler compounds were performed. For a large group of ferro- and antiferromagnetic Heusler alloys from the $X_2MnZ$ ($X$ = Co, Ni, Cu, Pb) series with the $L2_1$ structure, calculations have shown that a state close to HMF *($N\downarrow(E_F)$ is practically zero)* occurs in $Co_2MnZ$ systems with $Z$ = Al, Sn [43] and $Z$ = Ga, Si, Ge [68].

In [69], band calculations were performed for 54 ternary Heusler compounds $X_2YZ$, where *X* is a *3d* transition metal ($X$ = Mn, Fe, Co; $Y$ = Y, Zr, Nb, Mo, Tc, Ru, Rh, Pd, Ag and $Z$ = Al, Si). It has been shown that seven of them, namely, $Mn_2NbAl$, $Mn_2ZrSi$, $Mn_2RhSi$, $Co_2ZrAl$, $Co_2NbAl$, $Co_2YSi$, and $Co_2ZrSi$ are HMFs with 100% spin polarization and, for the other five alloys: $Mn_2TcAl$, $Mn_2RuAl$, $Mn_2NbSi$, $Mn_2RuSi$, and $Fe_2NbSi$, high spin polarization (more than 90%) with a gap for one of the spin directions near the Fermi level can be observed. These compounds were classified in [69] as "almost HMF," and the position of their Fermi levels $E_F$ is changed by applying pressure, so that the Fermi level falls into the gap, leading to the HMF state.

Calculations of the electronic structure and magnetic moment of half and full Heusler alloys $X$MnSb ($X$ = Ni, Pd, Pt, Co, Rh, Ir, Fe) and $Co_2MnZ$ ($Z$ = Al, Ga, Si, Ge, Sn), respectively, were presented in [70], where it was shown that the electronic and magnetic properties of these compounds are largely determined by the presence of a gap at the Fermi level for spin down electronic states. In this case, a linear dependence of the total magnetic moment *M* on the number of valence electrons *z* is observed: for half-Heusler alloys, $M=z - 18$ and, for full Heusler alloys, $M=z - 24$, which can be used to search and develop new HMF alloys with predetermined magnetic characteristics.

The results of calculating the electronic structure and phase stability of the Heusler alloys $Co_2YSi$ ($Y$= Ti, V, Cr, Mn, Fe, Co, Ni) are presented in [71]. The calculations have shown that the alloys $Co_2TiSi$, $Co_2VSi$, and $Co_2CrSi$ are HMF compounds. In this case, $Co_2CrSi$ has a high density of states at the Fermi level for the spin up states and a gap for the spin down states, as well as a high Curie temperature $T_C$ = 747 K. This can lead to 100% spin polarization of current carriers, even in the region of room temperatures. However, according to the calculations of [71], the HMF state in $Co_2CrSi$ is metastable. At the same time, the $Co_2TiSi$ and $Co_2VSi$ compounds are thermodynamically stable, although they exhibit a lower density of spin up states at $E_F$ and lower values of $T_C$.

It is known that atomic disorder can lead to a change in the type of structure, the appearance of additional structural phases, and, consequently, to a decrease in the spin polarization in HMF alloys. To elucidate the role of atomic disorder in the HMF state, in [72], first-principles calculations were performed for 25 compositions of cobalt-based Heusler alloys $Co_2YZ$ ($Y$ = Ti, V, Cr, Mn, Fe; $Z$ = Al, Ga, Si, Ge, Sn). It was shown that the disorder in the arrangement of the Co and $Y$ atoms correlates with the total charges of the valence electrons around the $Y$ atom, since the difference in the charges of the valence electrons between the Co and $Y$ atoms leads to different forms of the local potential at each site. This means that the compounds with titanium: $Co_2TiAl$, $Co_2TiGa$, $Co_2TiSi$, $Co_2TiGe$, and $Co_2TiSn$, are more convenient from a practical point of view for preventing atomic disorder in comparison with alloys $Co_2CrZ$, $Co_2MnZ$, and $Co_2FeZ$ ($Z$ = Al, Ga, Si, Ge, Sn).

Calculations of the electronic and magnetic properties of Heusler quaternary compounds with the chemical formula $XX'YZ$ (where $X, X'$, and $Y$ are transition metal atoms and $Z$ are $s$ and $p$ elements) with the structure of LiMgPdSn were performed in [73]. In this case, the valence of $X'$ atoms must be lower than the valence of the $X$ atoms and the valence of the element $Y$ is lower than the valence of both $X$ and $X'$. Sixty compounds $XX'YZ$ ($X$ = Co Fe, Mn, Cr; $X$ = V, Cr, Mn, Fe; $Y$ = Ti, V, Cr, Mn; $Z$ = Al, Si, As) were studied, and it was found that all of them obey the Slater – Pauling rule. In this case, 41 compounds are in the HMF state, 8 are in the SGS state, and 9 are semiconductors. In addition, CoVTiAl and CrVTiAl can be classified as ferromagnetic and antiferromagnetic semiconductors, respectively, with large energy gaps for both spin projections. It has been suggested that all HMF and SGS compounds will have high Curie temperatures, which makes them suitable for use in spintronics and magnetoelectronics.

Structural, electronic, and magnetic characteristics of 18 quaternary cobalt-based Heusler compounds $CoX'YSi$, where $X'$ = Y, Zr, Nb, Mo, Tc, Ru, Rh, Pd, and Ag and $Y$ = Fe and Mn were investigated using first-principles calculations in [74]. Several new half-metallic ferromagnets: CoNbMnSi, CoTcMnSi, CoRuMnSi, CoRhMnSi, CoZrFeSi, and CoRhFeSi, with 90 – 100% spin polarization were predicted, and 3 of them: CoRuMnSi, CoRhMnSi, and CoRhFeSi, can be used in spintronics at room temperature.

In [75], calculations and experimental studies of the CoFeCrAl alloy, which has a $B2$-type Heusler cubic structure, were performed. Calculations of the electronic structure predict the presence of an HMF state with a spin splitting energy of 0.31 eV, as well as a relatively high degree of spin polarization of carrier, *P*. An experiment using the Andreev spectroscopy of point contacts in the liquid helium temperature range gives the value of *P* = 0.67, which proves the validity of the calculations, and the linear dependence of the electrical resistivity on temperature indicates the possibility of sufficiently large *P* at higher temperatures.

According to calculations of the electronic structure and magnetic moment of the Heusler quaternary alloys CoFeCrP, CoFeCrAs, and CoFeCrSb [76], they are all half-metallic ferromagnets with gaps of 1.0, 0.52, and 0.75 eV, respectively. In this case, the total magnetic moment, to which the Cr atoms make the main contribution, is equal to $4\mu_B$ per formula unit and obeys the Slater – Pauling rule.

*4.4. Optical properties*

An interesting feature was discovered in experiments studying the optical properties of Heusler alloys. In conventional ferromagnetic Heusler alloys, the so-called Drude rise – an increase in optical conductivity $\sigma(\omega)$ with decreasing energy in the infrared (IR) region – is observed, whereas, in compounds exhibiting HMF properties, this rise is often absent [31-33].

In metals and alloys, there are two main mechanisms that determine their optical properties. In the IR region of the spectrum, the main role is played by the mechanism of intraband (i.b.) acceleration of electrons by the field of a light wave within one band [77]. Its contribution $\sigma_{i.b.}$ is determined by the parameters of the conduction electrons – the plasma frequency $\Omega$ and the relaxation frequency $\gamma$ – and decreases proportionally to the square of the frequency of the incident light, ro. It is described by the Drude formula:

$$\sigma_{i.b.} = \frac{1}{4\pi} \frac{\Omega^2 \cdot \gamma}{(\omega^2 + \gamma^2)} \tag{6}$$

In the visible and ultraviolet regions, the second mechanism dominates: interband (quantum) absorption of light with an electron transfer from a band below the Fermi level to a band above the Fermi level. The interband contribution is determined by the structure of the energy bands and carries information about the electronic states.

Since 1996, in many Heusler alloys, the absence of the Drude rise and a high level of interband absorption in the infrared region were found up to the boundary of the frequency range under study [31 – 33, 78, 79]. This behavior of the optical functions is anomalous for metallic systems and, as it turned out, correlates with the anomalous behavior of the electrical resistivity and, in some cases, agrees with the calculations of the electronic structure. Static conductivity at room temperature has low values, reaching which in the limit $\omega \rightarrow 0$ often implies a further decrease in optical conductivity. This anomalous behavior of the optical properties of alloys is apparently associated with the manifestation of an energy gap in the optical spectra for one of the spin projections.

Figure 10 shows the dispersion of the optical conductivity $\sigma(\omega)$ of Co and Fe based alloys in the energy range *E* = (0.1 – 2.0) eV [32]. It can be seen that the dispersion of $\sigma(\omega)$ of the Co$_2$TiAl and Fe$_2$TiAl alloys is typical of metals: a Drude rise is observed in the IR range and there are maxima due to interband absorption. For other alloys, anomalous behavior of optical conductivity in the IR range, i.e., high level of interband absorption and the absence of the Drude contribution, was found. The latter indicates low carrier concentrations, and, consequently, low values of conductivity. The static conductivity $\sigma_{st}$, which is the limit of optical conductivity at $\omega \to 0$, was also obtained from resistivity measurements; it is shown by symbols on the ordinate axis, the agreement is good enough (Figs. 10 and 11).

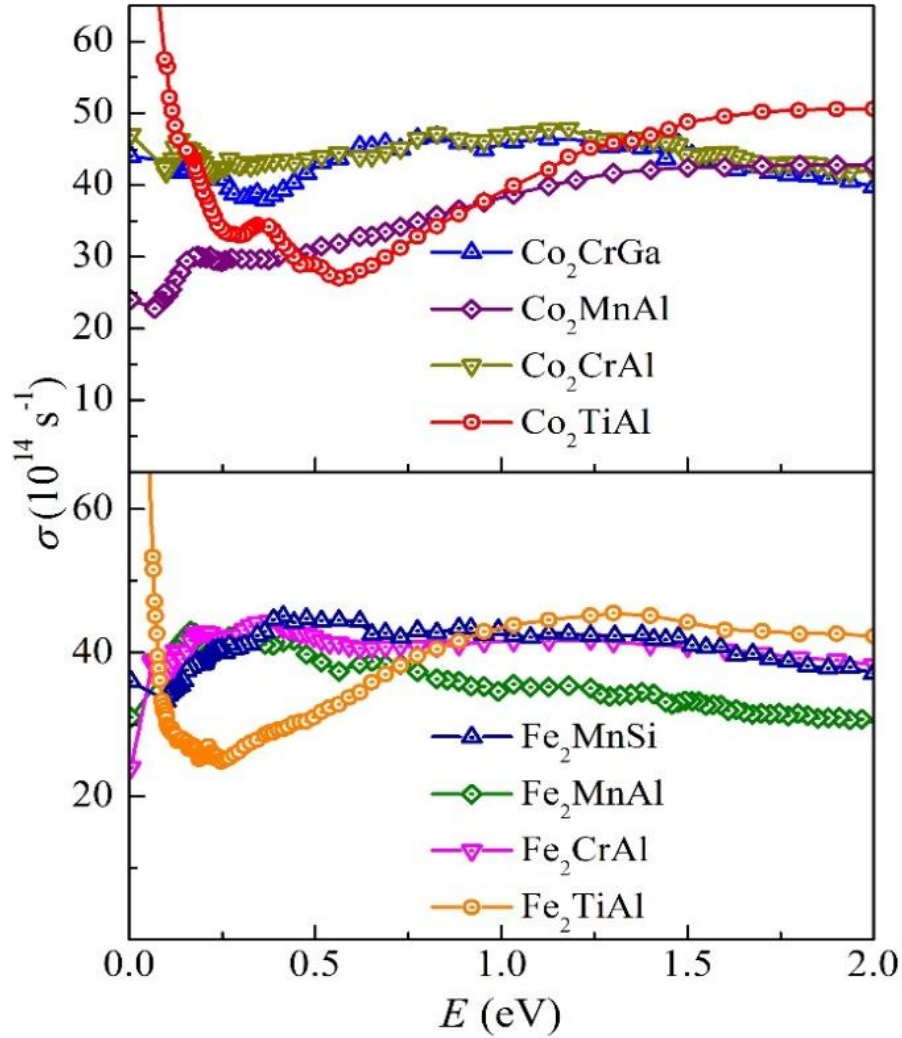

**Fig. 10.** Optical conductivity σ of Co- and Fe-based alloys, according to [32].

Figure 11 shows the temperature dependences of the resistivity $\rho(T)$. In the case of $Fe_2TiAl$ and $Co_2TiAl$, they are typical of metals with a residual resistivity of 12 and 40 μΩ×cm, respectively. For $Co_2CrGa$ and $Fe_2MnSi$, the resistivity increases with temperature, exhibiting metallic behavior. Nevertheless, their residual resistivity is much higher: 177 μΩ×cm for $Co_2CrGa$ and 154 μΩ×cm for $Fe_2MnSi$. The dependence $\rho(T)$ for $Co_2MnAl$ and $Fe_2MnAl$ is rather weak, while, for $Co_2CrAl$ and $Fe_2CrAl$, the resistivity decreases with increasing temperature (Fig. 11).

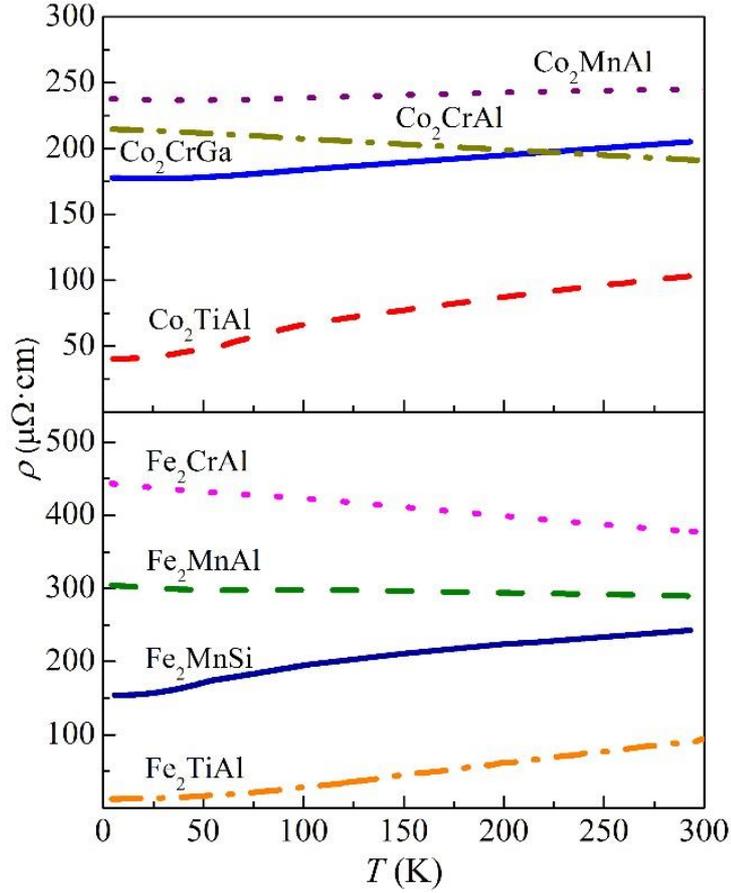

**Fig. 11.** Temperature dependences of electrical resistivity p of Co- and Fe-based alloys, according to [32].

In [32], the densities of states (DOS) were calculated for $Co_2TiAl$ and $Co_2CrAl$. Analysis of optical and calculated data, as well as electrical resistivity taking into account the existing band calculations [31, 80, 81] allowed the authors of [32] to conclude that these alloys have anomalous optical properties, which are qualitatively explained as follows. Such alloys have a high density of *d* states of *X* and *Y* atoms at the Fermi level and in its neighborhood for spin up electronic states. These electronic states make the main contribution to interband transitions in the IR range and also make a certain contribution to static conductivity. At temperatures below the gap, the probability of interband transitions for spin down electronic states is small; therefore, they practically do not contribute to the static conductivity.

## 5. Spin gapless semiconductors

Spin gapless semiconductors (SGSs) were predicted by Wang in 2008 [6]. The SGS materials have an unusual band structure: near the Fermi level, there is an energy gap for the subsystem of spin down electrons and, for spin up charge carriers, the top of the valence band touches the bottom of the conduction band. In such SGS materials, it is possible to realize (1) 100% spin polarization of charge carriers, (2) "separation" of spin-polarized electrons and holes using the Hall effect, (3) control of the Fermi level by means of an external electric field, and (4) control of the type (electrons or holes) and density of spin-polarized charge carriers. Spintronic devices require materials with a high degree of polarization of current carriers and a long spin relaxation time. Therefore, SGS compounds with the above-listed properties (1 – 4) may be one of the most suitable materials.

From a theoretical point of view, spin gapless semiconductors are close to the class of half-metallic ferromagnets. In SGSs, one can also expect nontrivial many-electron effects due to the interaction of electronic and spin degrees of freedom; however, the corresponding theoretical approaches have not yet been developed in detail.

In Heusler compounds, the SGS state is observed, as a rule, in the so-called inverse Heusler alloys with the $X_A$ structure. In [11], it was reported about one of the first experimental observations of the SGS state in $Mn_2CoAl$ with a Curie temperature of 720 K and a magnetic moment of $2\mu_B$/f.u. at 4.2 K. In the range from 4.2 to 300 K, its conductivity weakly depends on temperature and is relatively small, about 240 S/cm at room temperature; the carrier concentration is $\sim 10^{17}$ cm$^{-3}$; and the Seebeck coefficient is close to zero. Almost simultaneously with [11], calculations of the electronic band structure of quaternary alloys such as CoFeMnSi, CoFeCrAl, CoMnCrSi, CoFeVSi, and FeMnCrSb were performed, and the possibility of realizing the SGS state in them was predicted [83].

In [84], the electronic and magnetic properties of the Heusler alloys $Mn_2CoAl$ and $Mn_2CoGa$, initial and doped by replacing some of the Al and Ga atoms with Cr and Fe, were calculated. Since the initial alloy $Mn_2CoAl$ belongs to SGS compounds, and $Mn_2CoGa$ belongs to HMFs, it was interesting to trace the change in the electronic structure and magnetic state in these alloys upon alloying. It turned out that Fe and Cr additives destroy the SGS nature in $Mn_2CoAl$, giving rise to delocalized states in the "former" band gap for spin down states. At the same time, slight doping with $Mn_2CoGa$ leads to a more stable HMF state, which, however, is destroyed if the doping level exceeds a certain value. As for the magnetic properties, the total magnetic moments of the doped compounds are higher than those of the original ones.

In [85], using the density functional method, the band structure of Heusler quaternary alloys *XX'YZ (X, X',* and *Y* are transition metals, and *Z* = B, Al, Ga, In, Si, Ge, Sn, Pb, P, As, Sb, and Bi) was calculated in order to search for SGS states in them. Using the empirical rule for compounds with 21, 26, or 28 valence electrons, it has been shown that there are 12000 possible chemical compositions, of which only 70 are stable SGSs. The conclusion about the stability of these compounds was made on the basis of estimation of their thermodynamic, mechanical, and dynamic stability. It was shown that, among 70 stable SGS compounds, all four types of SGS I-IV states predicted in [6] can be realized. Type-II compounds can exhibit unique transport properties, in particular, anisotropic magnetoresistance and anomalous Nernst effect, which can be used in spintronic devices.

The results of studies of the structure and magnetic and electron transport properties of the cobalt-enriched Heusler quaternary SGS alloy $Co_{1+x}Fe_{1-x}CrGa$ (0 < *x* < 0.5), as well as calculations of the band structure, are presented in [86]. A specific feature of these alloys is a high Curie temperature and magnetization, which vary from 690 K ($x = 0$) to 870 K ($x = 0.5$) and from 2.1 $\mu_B$/f.u. ($x = 0$) up to 2.5 $\mu_B$/f.u. ($x = 0.5$), respectively. Analysis of the temperature dependence of the electrical resistivity shows that the alloys exhibit SGS properties up to $x = 0.4$ and, at $x = 0.5$, they exhibit metallic behavior both in magnitude and in the form of the temperature dependence. Unlike conventional semiconductors, the conductivity (resistivity) in these SGS compounds at 300 K lies in the range from 2300 to 3300 S/cm (from 300 to 430 $\mu\Omega\times$cm), which is close to those of other SGS materials; anomalous Hall resistivity increases from 38 S/cm for $x = 0.1$ to 43 S/cm for $x = 0.3$. The Seebeck coefficient turns out to be vanishingly small below 300 K, which is another indicator of the realization of the SGS state.

In [87], the structure, electron transport, and magnetic properties were experimentally studied and the electronic structure of the quaternary CoFeCrGa alloy with the structure $L2_1$, in which chemical disorder is observed, was calculated. For the saturation magnetization at $T = 8$ K, the Slater – Pauling rule is satisfied, and the Curie temperature exceeds 400 K. The high resistivity, low carrier concentration, and weak temperature dependence also indicate the realization of the SGS state. It was shown that, under pressure, the SGS compound CoFeCrGa can transform into the HMF state, which is explained by the peculiarities of its electronic structure.

The magnetic and galvanomagnetic properties of thin films of the Heusler alloy $Mn_{2-x}Co_{1+x}Al$ ($0 \leq x \leq 1.75$) were experimentally studied in [88]. From measurements of the magnetization, it follows that, in films with $x = 1.75, 1.5, 1.25$, and 1, $Mn_{2-x}Co_{1+x}Al$ ($0 \leq x \leq 1.75$), one observes a ferromagnetic order and, with $x = 0, 0.5$, and 0.75, a ferrimagnetic order. In $Mn_2CoAl$ films, a semiconducting behavior of conductivity is observed. Together with low values of the anomalous Hall conductivity (about 3.4 S/cm at 10 K) and a positive magnetoresistivity linear in the field, this may indicate an SGS state, especially in films with a high Mn concentration (small values of $x$). Based on the results obtained, the authors of [88] conclude that such films can be useful in semiconductor spintronics.

Experimental studies of the structure, magnetic properties, electron transport, and Andreev reflection and calculations of the band structure of the Heusler quaternary equiatomic alloy CoFeMnSi were performed in [89]. This alloy crystallizes in a cubic Heusler structure (of the LiMgPdSn type) and has a Curie temperature of 620 K and a saturation magnetization of 3.7$\mu_B$/f.u., the behavior of which obeys the Slater – Pauling rule. The latter, as noted in [89], is one of the necessary conditions for the realization of an SGS state. Low values of electrical conductivity (about 3000 S/cm at $T = 300$ K) and carrier concentration (about $5 \times 10^{19}$ cm$^{-3}$ at $T = 300$ K) and their weak temperature dependences are interpreted as the appearance of a state of a spin gapless semiconductor. The anomalous Hall conductivity in this case is 162 S/cm at 5 K. Using the Andreev reflection technique, the spin polarization of the current carriers was found to be $P = 0.64$.

In [90], where first-principles calculations of the electronic structure and magnetic properties of the Heusler alloys $Ti_2CoSi$, $Ti_2MnAl$, and $Ti_2VAs$ were performed, it was shown that their magnetic state strongly depends on the structure. If these compounds are ordered in a $X_A$ structure, then they exhibit SGS properties. In this case, the $Ti_2MnAl$ and $Ti_2VAs$ alloys have zero magnetization, which may indicate a state of compensated ferrimagnetism. If all these alloys are ordered in an $L2_1$ cubic structure, they are metals.

A brief review of works concerning the calculation of the electronic structure of inverse Heusler ternary compounds $X_2YZ$ (where $X$ and $Y$ are transition metal atoms, with $X$ having the lowest valence) and quaternary Heusler alloys $XX'YZ$ (with a structure of LiMgPdSn) is presented in [91]. Depending on the chemical composition, due to the existence of three magnetic sublattices, ferro-, ferri-, or antiferromagnetism can arise in them. The most promising for practical application are compounds with 18 valence electrons, since the total magnetization in them is zero, which leads to minimal energy losses. Some of these alloys are spin gapless semiconductors, and the most interesting of them is $Mn_2CoAl$, which, apparently, can be used in spintronics.

In [92], the optical properties of the $Mn_{1.8}Co_{1.2}Al$ alloy, which is close in composition to the spin gapless semiconductor $Mn_2CoAl$, were investigated and its electronic structure was calculated. Anomalous behavior of the optical properties of this alloy was found. The positive values of the real part of the permittivity $\varepsilon_1$ and the absence of the Drude contribution to the optical conductivity in the IR spectral region up to the boundary of the interval under study indicate a weakening of the metallic properties of the alloy. Intense interband absorption in the IR region indicates a complex structure of the band spectrum and a high density of the $d$ states in the vicinity of $E_F$. These features of the optical absorption spectrum make it possible to explain the band spectrum pattern that is characteristic of spin gapless semiconductors.

In [93], the microstructure and the magnetic and transport properties of the Heusler alloy $Mn_2CoAl$ were investigated. Microstructural analysis showed that the main phase was not $Mn_2CoAl$, but $Mn_{1.8}Co_{1.4}Al_{0.8}$, which has a disordered inverse $X_A$ structure with the positions of Mn (A) and Al(C) atoms partially substituted by Co and Mn atoms, respectively. Calculations of the electronic structure of this phase showed that it corresponds to the HMF state, in contrast to the SGS phase of $Mn_2CoAl$. At the same time, large values of the electrical resistivity of the studied samples, the "semiconductor" form of its temperature dependences, and low values of the anomalous Hall conductivity are characteristic features of the SGS state. As stated in [93], this may be due to the localization of charge carriers due to the crossing of the Fermi level by the $3d$ states of Mn and $3d$ states of Co in $Mn_{1.8}Co_{1.4}Al_{0.8}$.

The structure and the magnetic and transport properties of quaternary alloy CoFeMnSi were studied in detail experimentally in [94]. The saturation magnetization of this compound is $3.49\mu_B$/f.u., and the Curie temperature is $T_C = 763$ K. Large values of the electrical resistivity, which decreases with increasing temperature, and the practically temperature-independent carrier concentration of $\sim 10^{20}$ cm$^{-3}$, may indicate the realization in the alloy CoFeMnSi of the SGS state.

## 6. Topological semimetals

Figure 12 schematically shows the band structures of various topological structures: topological insulators and topological Dirac and Weyl semimetals.

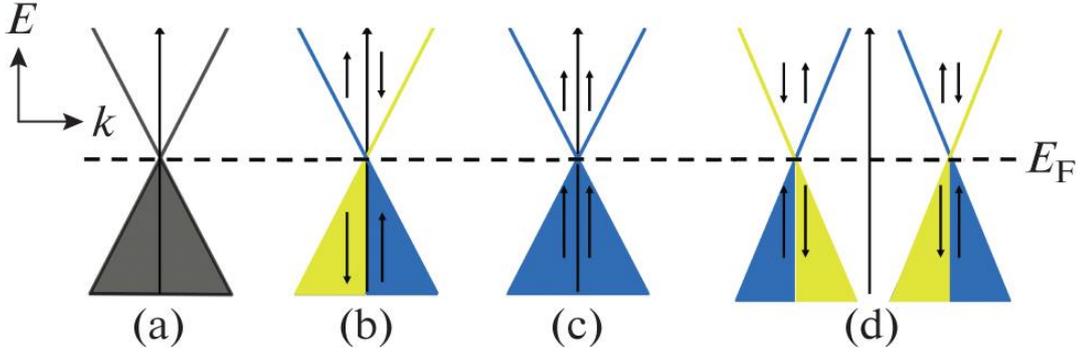

**Fig. 12.** Band structures of topological systems: (a) spin-degenerate systems, (b) surface states of topological insulators, (c) topological Dirac states, and (d) Weyl semimetals. The arrows show the spin directions.

A nontrivial topology of the electronic band structure arises in materials with strong spin-orbit interaction, which leads to inversion of the conduction and valence bands. Topological insulators [95-98] opened up a class of materials with nontrivial topological properties, which later also included topological semimetals [95, 96, 99, 100].

Topological insulators (TI) are a class of narrow-gap materials in the bulk of which there is a characteristic energy gap and, on the surface, there are "metallic" states. Gapless surface states arise from band inversion due to strong spin-orbit interaction and are protected by time reversal symmetry. The electrons in them are Dirac fermions with a linear dispersion law, and their spins are rigidly related to the momentum. TI states have been observed in many materials, including HgTe with quantum wells [101], $Bi_2Se_3$, $Bi_2Te_3$, $Sb_2Te_3$ [102], and some Heusler alloys [103, 104].

Topological Weyl and Dirac semimetals generalize the classification of topological materials. In materials such as TIs, the band structure in the bulk has a gap, arising due to the strong spin-orbit interaction, with the exception of the intersection points of the bands at the Weyl or Dirac nodes. Near these nodes, the band dispersion in all three directions of momentum space is linear and low-energy excitations can be described by the Weyl or Dirac Hamiltonians.

Weyl semimetals are characterized by the presence in their volume of the so-called Weyl fermions, which have zero effective mass. The existence of a massless fermion was first predicted in 1929 by Hermann Weyl. However, such quasi-particles were detected experimentally only relatively recently in the TaAs family [100]; later, in $T_d$-$MoTe_2$ [105] and $WTe_2$ [106], and in magnetic materials based on Heusler alloys [9, 107, 108]. Weyl nodes always appear in pairs of opposite chirality; they can be regarded as monopoles of Berry curvature in momentum space. A consequence of the nontrivial topology of the bulk band structure in Weyl semimetals are unique topologically protected surface states: Fermi arcs, which connect the projections of Weyl nodes onto the surface. In contrast to Weyl semimetals, Dirac semimetals do not have chirality: a Dirac node can be considered as the "sum" of two Weyl nodes of opposite chirality. Dirac semimetals include, e.g., $Cd_3As_2$ [109] and $Na_3Bi$ [110]. The difference in the electronic band structure of Dirac and Weyl semimetals is due to symmetry. The condition for the existence of a Weyl semimetal is the violation of symmetry with respect to the inversion or time reversal. The interplay of both of these factors in the crystal leads to the appearance of the Dirac semimetal phase. During a phase transition from a TI to a normal insulator, the Dirac or Weyl phase can be obtained as an intermediate phase, which depends on whether the inversion symmetry is violated or not [111].

Weyl nodes can lead to unusual transport properties, including the giant anomalous Hall effect (AHE). The corresponding (internal) contribution to this effect is calculated by integrating the Berry curvature over the entire Brillouin zone, which is the equivalent of a magnetic field in momentum space. In particular, for a Weyl semimetal with two Weyl nodes, the anomalous Hall conductivity is proportional to the distance between these nodes [9]. From this point of view, the AHE in three-dimensional Weyl semimetals [9, 99] can be related to the quantum AHE in a two-dimensional situation (see also the discussion in [112]).

A large AHE was discovered in the half-Heusler antiferromagnet GdPtBi [113]. The authors of [114] managed to grow a bulk single crystal of the TSM alloy $Co_2MnGa$ and to study its electronic structure and magnetic state using ARPES spectroscopy, the density functional method, and measurements of electrical resistivity, Hall effect, and magnetization. As a result, the emergence of a topological phase in the $Co_2MnGa$ alloy was found: the lines of topological nodes intersect the Fermi level and surface states arise, located along the lines of nodes in the Brillouin zone near the surface. Experimental studies of the volume and surface of a $Co_2MnGa$ single crystal lead to a conclusion about the realization of the first topological 3D magnet and, according to the authors of [114], the first topological magnetic metal. The presence of lines of topological nodes can explain the very large anomalous Hall conductivity observed in $Co_2MnGa$. The detection of them allowed the authors to propose a method for creating strongly spin-polarized current carriers.

In $Mn_3Ge$ and $Mn_3Sn$ compounds with a distorted Heusler structure $D0_{19}$, in the *ab* plane, a triangular lattice of Mn atom is formed – a highly frustrated kagome lattice [115, 116] (Fig. 13).

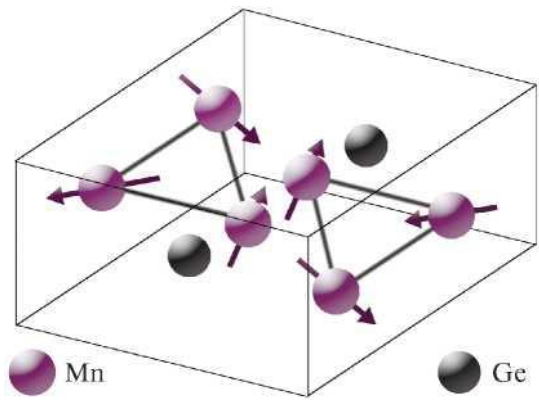

**Fig. 13.** $D0_{19}$ structure of $Mn_3Ge$ (according to [116]). The arrows show the direction of the magnetic moments of the Mn atoms.

In chiral antiferromagnetic compounds $Mn_3X$ *(X = Ge, Sn, Ga, Ir, Rh, Pt)*, a strong anisotropic anomalous Hall effect and a spin Hall effect were found [117]. Note that the $Co_3Sn_2S_2$ compound with a kagome lattice and nontrivial topological properties is also a representative of half-metallic ferromagnets [112, 118].

Various Heusler compounds based on heavy elements with strong spin-orbit interaction (SOI), their electronic structure, and properties are considered in the review [9]. It is shown that the combination of the features of symmetry, details of the SOI, and the magnetic structure makes it possible to realize, due to the emerging Berry curvature, a wide range of topological phases, including topological insulators and semimetals. In such compounds, Weyl points and lines of nodes can appear, which can be controlled using various external actions. This leads to a number of exotic properties such as chirality anomalies, large anomalous spin, and topological Hall effects. Due to the noncollinear magnetic structure and Berry curvature, a nonzero anomalous Hall effect can arise, which was first observed in the antiferromagnets $Mn_3Sn$ and $Mn_3Ge$. Along with the peculiarities of the Berry curvature in *k*-space, in Heusler compounds with non- collinear magnetic structures, topological states in the form of magnetic antiskyrmions in real space can also arise. In the review [9], it was told about a direct effect on the Berry curvature and, consequently, the possibility of purposeful control of the electronic and magnetic structure and properties of such compounds.

In [107], a new family of Weyl systems based on Heusler alloys $Co_2Y$Sn *(Y = Ti, V, Zr, Nb, Hf)* was predicted. In such centrosymmetric materials, there are only two Weyl nodes at the Fermi level (the minimum possible number), which are protected by rotational symmetry along the magnetic axis and are separated by a large distance in the Brillouin zone. In this work, the corresponding Fermi arcs were also calculated. The results obtained can contribute to the understanding of magnetic effects in Weyl semimetals.

Calculations of the electronic structure for HMF Heusler alloys in order to study the stability of the Weyl points were performed in [119] for $Co_2TiAl$. The number of Weyl points and their coordinates in *k*-space were controlled using the orientation of the magnetization. This alternative degree of freedom, which is absent in other topological materials (e.g., in ordinary Weyl semimetals), suggests new functional characteristic of half-metallic ferromagnets.

In [120], the electronic structure of the inverse Heusler alloy $Ti_2MnAl$ was calculated and the realization of the state of a Weyl magnetic semimetal in it was predicted. The $Ti_2MnAl$ alloy was chosen because it is a compensated ferrimagnet with a Curie temperature above 650 K and a carrier concentration of about $2 \times 10^{19}$ cm$^{-3}$. Despite the zero total magnetic moment, it exhibits a large intrinsic anomalous Hall effect (AHE) with a corresponding conductivity of about 300 S/cm, which is a consequence of the redistribution of the Berry curvature from the Weyl points, which are located only 14 meV from the Fermi level and are isolated from trivial bands. It was shown in [120] that, in contrast to the antiferromagnets $Mn_3X$ *(X = Ge, Sn, Ga, Ir, Rh, and Pt)*, where the AHE appears due to the noncollinear magnetic structure, the AHE in $Ti_2MnAl$ arises directly from the Weyl points and is topologically protected. Compared to Co-based ferromagnetic Heusler compounds, due to the absence of mirror symmetry in the inverse Heusler structure, the Weyl nodes in $Ti_2MnAl$ do not arise from the node lines. Since the magnetic structure violates rotational symmetry, the Weyl nodes are stable even in the absence of spin-orbit interaction. This is one of the first examples of a TSM material with Weyl points and a large anomalous Hall effect, where the total magnetic moment is zero.

Calculations of the electronic structure, experimental studies of the magnetic, electrical, and thermoelectric properties of single-crystal ferromagnetic TSM $Co_2MnGa$ were performed in [121]. A large value of anomalous Nernst thermopower of about 6 μV/K was found in a field of 1 T at room temperature, which is about 7 times higher than any value ever observed for an ordinary ferromagnet. Such a high value of anomalous Nernst effect (ANE) arises due to the large total Berry curvature near the Fermi level, associated with the node lines and Weyl points, although the AHE is usually proportional to the magnitude of magnetization.

It is known that some of cobalt-based Heusler compounds, to which, in particular, the $Co_2TiSn$ alloy belongs, are half-metallic ferromagnets with Weyl points. In the absence of T-symmetry with respect to time reversal, these systems have a Berry curvature in momentum space, which can lead to anomalies in electron transport. For this purpose, the electronic structure was calculated and the electro- and magnetoresistivity, the Hall effect, and magnetization were measured on epitaxial $Co_2TiSn$ films [122]. Theoretical analysis showed that the Berry curvature plays an important role in the formation of the anomalous Hall effect and the resulting discrepancy between theory and experiment is associated with the mechanisms of side jumps and skew scattering. It was shown in [122] that the intrinsic contribution to the anomalous Hall effect arises from the node lines, partially broken due to the lowering of symmetry caused by the absence of the *T*-symmetry.

In computational work [123], the role of chemical disorder in the evolution of the Weyl points in magnetic TSM $Co_2Ti_{1-x}V_xSn$ ($0 \leq x \leq 1$) was studied. Calculations of the electronic structure, magnetic moment, and anomalous Hall conductivity were performed using the first-principles approach to track the evolution of Weyl node lines. As the vanadium concentration increases to $x = 0.5$, the node line moves to the Fermi level. The calculation of the density of states shows the TSM behavior for all compositions investigated. With increasing V concentration, the magnetic moment on the Co atom increases to a maximum at $x = 0.4$ and then begins to decrease, while the total moment permanently increases. Calculations show that, when almost half of Ti is replaced by V, the intrinsic anomalous Hall conductivity increases almost two-fold in comparison with the undoped composition. It is concluded that a composition close to $Co_2Ti_{0.5}V_{0.5}Sn$ must be ideal for the experimental study of Weyl TSMs.

Experiments on measuring the inverse spin Hall effect in Heusler compound $Co_2MnGa$, which is a ferromagnetic Weyl TSM, were carried out at room temperature using the method of spin injection in "spin valve" structures [124]. The value of the angle $\theta_{SH}$ of the spin Hall effect turned out to be one of the highest for ferromagnets: $\theta_{SH} = -0.19 \pm 0.04$. In addition, in this case, the Onsager principle is not satisfied, which can partly be explained by different values of the Hall conductivity for spin up and down carriers.

To obtain new information on the nontrivial topology of the Bloch states, which leads to anomalous contributions to the transport coefficients, it was proposed in [125] to study the off-diagonal components of the corresponding tensors of conductivity and thermal conductivity. The object of study was also Weyl ferromagnetic TSM $Co_2MnGa$, and its magnetic, electrical, galvanomagnetic, thermal, and thermoelectric properties were measured. It was shown that the Wiedemann-Franz law, which relates the electrical and thermal coefficients, is fulfilled in a wide temperature range from helium temperatures to room temperatures. The authors of [125] assert that the anomalous Hall effect has an internal origin and the Berry spectrum near the Fermi level has no singularities and is smooth. From the data on the anomalous Nernst effect, the magnitude and the temperature dependence of the anomalous thermoelectric conductivity $α_{ij}^A$ was obtained and it was shown that its ratio to the anomalous Hall conductivity $α_{ij}^A$ approaches the ratio of constants $k_B/e$ at room temperature.

Topological features can also manifest themselves in the magnetic state of TSM materials, leading to unusual magnetic characteristics, in particular, to noncollinear magnetic modulation of the Weyl nodes. In [126], where the electronic structure and magnetic properties of the tetragonal ferrimagnetic compound $Mn_3Ga$ were calculated, it was reported that topologically nontrivial node lines appear in the absence of spin – orbit interaction (SOI), which are protected by both mirror and rotational $C_{4z}$ symmetry. It was shown that, in the presence of SOI, double-degenerate nontrivial intersection points evolve into $C_{4z}$-protected Weyl nodes with a chiral charge of ±2. Double Weyl nodes are divided into a pair of Weyl nodes with a charge of – 1, which is determined by the magnetic orientation in a noncollinear ferrimagnetic structure.

It was also reported in [127] that the magnetization of a topological Weyl semimetal can lead to the appearance of anomalous transport properties due to its topological nature. The authors of [127] used ARPES spectroscopy and calculations of the electronic structure and also measured the electron transport and magnetic properties of $Co_2MnGa$ ferromagnetic films. They managed to visualize the spin-polarized Weyl cone and observe dispersionless surface states. It was found that, in the range of room temperatures, the anomalous Hall and Nernst conductivities increase as the magnetization-induced massive Weyl cone approaches the Fermi level until the anomalous Nernst thermoelectric power reaches 6.2 $μ_B/K$. The discovered relationship between the topological quantum state and remanent magnetization made it possible to conclude that $Co_2MnGa$ films can be used to create highly efficient devices for measuring heat flux and magnetic field, operating at room temperature and in zero magnetic field.

The band structure of Heusler alloy $Co_2MnGe$ was studied in [128] using ARPES spectroscopy, and the data obtained were compared with the results of first-principles calculations. A broad band with parabolic dispersion at the center of the Brillouin zone and several main spin bands with high spin polarization, crossing the Fermi level near its boundary, were identified. The ARPES spectroscopy data confirm the absence of the contribution of minority spin bands on the Fermi surface, which means the appearance of the TSM state. In addition, two topological Weyl cones with points of intersection of the bands near the point X were identified. The authors of [128] conclude that, in $Co_2MnGe$, a TSM state arises, which should lead to large anomalous Hall and Nernst effects.

In [129], a systematic analysis of experimental data on the anomalous Hall effect (AHE)

in $Co_2TiZ$ (Z = Si and Ge) was presented in order to understand the role of the Berry curvature in the formation of the AHE. It was found that the anomalous Hall resistivity changes depending on the electrical resistivity according to a law close to quadratic for both compounds. A detailed analysis shows that the contribution to the anomalous Hall conductivity from the internal Karplus—Luttinger mechanism associated with the Berry phase prevails over the external contributions from asymmetric scattering and side-jump mechanisms.

Thus, the study of topological phenomena in Heusler alloys is just beginning and seems very promising: here, in combination with a large spontaneous magnetization, a number of physically interesting and practically important effects are possible.

## 7. Evolution of states in Heusler alloys

In practice, it is not easy to rigorously realize the states of a half-metallic ferromagnet, a spin gapless semiconductor, and a topological semimetal; therefore, when interpreting experiments, ambiguous situations often arise. In this case, the features of the electronic structure (the density of electronic states near the Fermi level) and, consequently, of the physical properties change very strongly upon varying the *Y*- and *Z*-components in Heusler compounds $X_2YZ$ (see, e.g., [59 – 63, 130 – 140]). In such cases, transitions can be observed from the ordinary (magnetic and nonmagnetic) metallic and semiconducting states to the state of a half-metallic ferromagnet, then to the state of a spin gapless semiconductor, topological semimetal, and vice versa. Since *Y* are usually *3d* transition metals and *Z* are elements of the III – V groups of the Periodic table, changes in the density of electronic states near the Fermi level and, consequently, in physical properties manifest themselves differently when the components are changed.

As a result of studying the electron transport and magnetic properties in alloys $X_2YZ$ (X = Mn, Fe, Co; Y = Ti, V, Cr, Mn, Fe, Co, Ni; Z = Al, Si, Ga, Ge, In, Sn, Sb), it was found that, by changing the *Y*- and *Z*- components, the systems can vary from ordinary metals to half-metallic systems and, possibly, spin gapless semiconductors [59 – 63, 137 – 140].

In particular, in alloys of the $Co_2FeZ$ system, with a change in the *Z*-component, small values of the residual resistivity $\rho_0$ and a metallic type of temperature dependences $\rho(T)$ are observed [58, 61, 137]. In this case, all alloys are ferromagnets with relatively high values of magnetization and Curie temperatures and the anomalous Hall coefficient $R_S$, as a rule, exceeds the normal Hall coefficient $R_0$ by several orders of magnitude. The carrier concentrations are typical of metals; the contribution to the resistivity is proportional to $T^n$, $7/2 < n < 9/2$, which may indicate two-magnon scattering processes. Thus, in addition to the properties of conventional ferromagnets, the properties of HMF are also manifested.

In $Fe_2YAl$ alloys [59], with a change in the *Y*-component, both relatively small ($Fe_2TiAl$, $Fe_3Al$, $Fe_2NiAl$) and huge ($Fe_2VAl$) values of the residual resistivity were observed, along with the metallic behavior of $\rho(T)$ and the presence of wide temperature regions with negative temperature coefficient of resistivity (TCR). The magnetic state of alloys varies from paramagnetism or weak itinerant magnetism ($Fe_2TiAl$, $Fe_2VAl$) to "good" ferromagnetism ($Fe_3Al$, $Fe_2NiAl$); the carrier concentrations vary from semiconductor values ($Fe_2VAl$) to values typical of "bad" metals ($Fe_2TiAl$, $Fe_3Al$, $Fe_2NiAl$). In addition to the properties of an ordinary ferromagnetic metal, the properties of a half-metallic ferromagnet are also manifested. The $Fe_2VAl$ alloy exhibits almost semiconducting properties. According to [18, 20], this system demonstrates a transition from a semiconductor to a metallic state with the appearance of a ferromagnetic order. Thus, in alloys of the $Fe_2YAl$ system, the features of the electron energy spectrum near the Fermi level play an important role in electron transport.

In the $Mn_2YAl$ alloys [62, 138], with a change in the $Y$-component, a relatively large residual resistivity $\rho_0$ is observed, except the $Mn_2VAl$ and $Mn_2NiAl$ alloys, for which the metallic behavior of $\rho(T)$ is observed. For other compounds of this system, there are regions with a negative temperature coefficient of resistivity. In contrast to calculations of the electronic structure [141], magnetic measurements for the $Mn_2CrAl$, $Mn_3Al$, and $Mn_2FeAl$ systems give zero total magnetization, which may indicate compensated ferrimagnetism.

Figure 14 (according to the data of [139, 140]) demonstrates correlations between the residual resistivity $\rho_0$, saturation magnetization $M_s$, and the coefficient of spin polarization of charge carriers, $P$, for the $Co_2YSi$ system ($Y$ = Ti, V, Cr, Mn, Fe, Co, Ni), i.e., when the number of valence electrons $z$ changes from 26 for $Co_2TiSi$ to 32 for $Co_2NiSi$. It can be seen that $\rho_0$ increases near $z = 28$, while a minimum is observed for the magnetization $M_s$ at this point. According to [139, 140], in the $Co_2MnSi$ and $Co_2FeSi$ alloys, HMF and SGS states can arise with a high degree of polarization of the charge carriers. As can be seen from Fig. 14, it is for these alloys that the maximum values of magnetization and minimum values of residual resistivity are observed. In [139, 140], this is explained by the fact that these compounds predominantly contain metallic spin up charge carriers, providing a metallic type of conductivity and a large magnetic moment, which ultimately must lead to a high spin polarization $P$ of charge carriers. It can be seen from the dependences presented in the figure [140] that, for the $Co_2MnSi$ and $Co_2FeSi$ alloys, rather high values of $P$ are observed, indeed.

The anomalies revealed may indicate the features of the energy spectrum of electrons: the appearance of states of a half-metallic ferromagnet or a spin gapless semiconductor.

In [144], it was reported about a zero magnetic moment in thin films of the Heusler alloy $Mn_3Al$, which was explained by the state of a compensated ferrimagnet (CFIM). The difference of this state from the antiferromagnetic (AFM) state is that the crystallographic positions of manganese are different. In [62], a zero magnetic moment was also observed in the cast $Mn_3Al$ alloy, and it was suggested that this may be a manifestation of compensated ferrimagnetism.

The observation of the CFIM state was reported in [145], where the magnetization and electrical and magnetic resistivity of Heusler alloys $Mn_{2-x}Ru_{1+x}Ga$ ($x$ = 0.2, 0.5) were experimentally studied. It was shown that the $Mn_{1.5}Ru_{1.5}Ga$ alloy exhibits a vanishingly small magnetic moment at low temperatures, which can be explained by the antiparallel arrangement of the moments of Mn atoms in two nonequivalent nodes, i.e., implementation of the CFIM state.

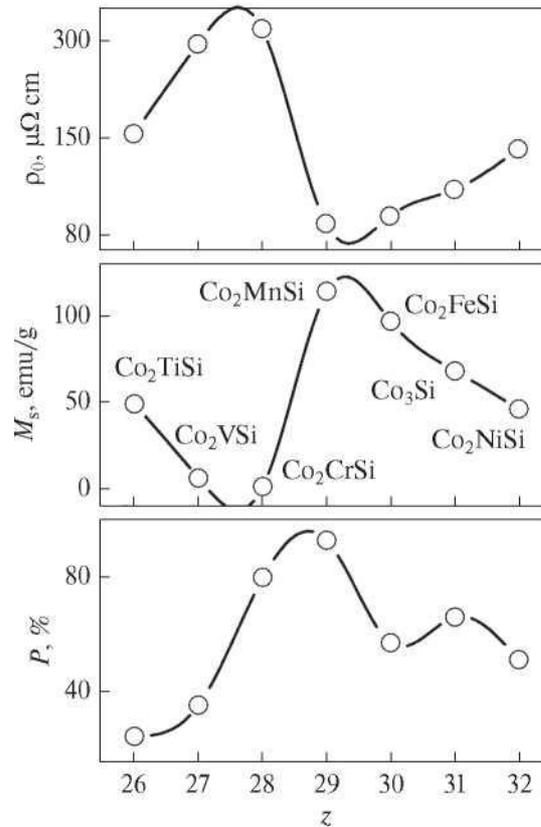

**Fig. 14.** Dependences of the residual resistivity $\rho_0$ and the saturation magnetization $M_s$, determined experimentally at $T = 4.2$ K and dependence of the polarization of the charge carriers $P$ for the Co$_2Y$Si alloys. The values of the coefficient $P$ are taken from [35, 71, 142, 143].

Experimental work [146] presents the results of band calculations and experimental data for the optical properties of the Co$_2$Cr$_{1-x}$Fe$_x$Al alloy, i.e., on the transition from ordinary ferromagnet Co$_2$FeAl to half-metallic ferromagnetic alloy Co$_2$CrAl. It was shown that the behavior of the optical properties in the Co$_2$FeAl alloy is typical of metals: there is a region of Drude absorption on the $\sigma(\omega)$ curve in the IR spectral region and a main absorption band in the visible and ultraviolet regions. In the Co$_2$Cr$_{0.4}$Fe$_{0.6}$Al, Co$_2$Cr$_{0.6}$Fe$_{0.4}$Al, and Co$_2$CrAl alloys, the behavior of the optical properties is anomalous: there is no section of the Drude rise in the $\sigma(\omega)$ curve in the IR region. The different behavior of the optical properties confirms the substantially different character of the electronic states in the system of bands with spins along the direction of magnetization, especially near the Fermi level. The position of the latter in the region of high density of the $d$ states for Co$_2$Cr$_{0.4}$Fe$_{0.6}$Al, Co$_2$Cr$_{0.6}$Fe$_{0.4}$Al, and Co$_2$CrAl causes a high level of interband absorption in the IR region, low values of the effective carrier concentration, and weak intraband absorption. On the contrary, in the Co$_2$FeAl alloy, those are the $sp$ states that reach the Fermi level, which causes a high level of intraband absorption, large values of static conductivity, and a weak contribution from interband absorption in the IR region.

The possibility of switching between two states – HMF and ordinary metal – was pointed out in computational work [147], where the electronic structure was calculated and the phase stability of ferromagnetic Heusler alloy Mn$_2$ScSi was analyzed. It was shown that two phases can exist in the alloy: the HMF phase, which occupies a small volume in the crystal lattice in a weak field, and a metallic FM phase of a large volume. If the lattice remains cubic, transitions can occur between these two states, caused by triaxial compression and expansion, as well as by an external magnetic field.

The electrical resistivity and galvanomagnetic properties of single crystals of the half-Heusler compound LuPtSb were experimentally studied in [148]. The magnitude of the electrical resistivity, the form of its temperature dependence, as well as the values of the carrier concentration, indicate that, at temperatures below 150 K, the LuPtSb compound is a gapless $p$-type semiconductor. This is also evidenced by the weakly temperature-dependent positive magnetoresistivity, reaching 109%, and the carrier mobility up to 2950 cm$^2$/V s. In addition, due to the strong spin-orbit coupling at temperatures below 150 K, weak antilocalization effects arise. At higher temperatures (>150 K), a transition from the semiconducting to the metallic state occurs.

The work [149] is devoted to the search and experimental study of the topological Hall effect in thin films of the Heusler SGS alloy Mn$_2$CoAl coated with a thin Pd layer. Using the strong dependence of the anomalous Hall effect on thickness and temperature, the authors managed to detect and study in detail the topological Hall effect at temperatures from 3 to 280 K. The presence of this effect indicates the existence of skyrmions: topologically nontrivial noncoplanar spin textures. Using a new method that takes into account the field loops of Hall resistivity, the coexistence of skyrmions of opposite polarity was demonstrated.

In the review [9], Heusler compounds (including spin gapless semiconductors and magnetically compensated ferrimagnets) and their properties are considered from a topological point of view. The relationship between topology and symmetry properties, non-collinear order in ferromagnetic and antiferromagnetic compounds, the anomalous Hall effect, and magnetic antiskirmions are discussed.

Table 1 presents the characteristics of various systems of full-Heusler alloys, which exhibit the properties of ordinary magnets, half-metallic ferromagnets, and spin gapless semiconductors. According to the available literature data, the table gives their ferromagnetic (FM), half-metallic ferromagnetic (HMF), or spin gapless semiconductor (SGS) state, as well as the residual resistivity, saturation magnetization, and the value of spin polarization.

The above examples clearly demonstrate the variety of properties of systems based on Heusler alloys, which reflect a number of actively developing areas of modern physics of condensed matter and quantum magnetism.

## 8. Conclusions

As we have seen, Heusler alloys, discovered more than 100 years ago and comprising about 1500 different compounds, have many unusual characteristics and functional properties. Of particular importance is the possibility of realizing the states of a half-metallic ferromagnet (HMF), a spin gapless semiconductor (SGS), and a topological semimetal

(TSM).

Half-metallic ferromagnetic Heusler alloys are characterized by the presence of a gap at the Fermi level for electronic spin down states with and its absence for spin up carriers. The behavior of electron transport in such systems can be described in a model with two conduction channels. Then, at temperatures much below the gap, the conductivity is mainly determined by one conduction channel, i.e., spin up charge carriers, since the spin down carriers are frozen out. In alloys with high Curie temperatures, e.g., $Co_2FeSi$ and $Co_2MnSi$, for which $T_C$ reaches 1100 K, even at room temperature only one conduction channel is working and the degree of polarization of current carriers is high. At the same time, non-quasiparticle states with an opposite spin projection, which have low mobility, can contribute to the spectral, magnetic, and thermodynamic properties and indirectly manifest themselves in scattering processes. They are especially important when considering the problem of spin polarization.

Spin gapless semiconductors (SGSs) are new quantum materials with a unique spin-polarized band structure. Unlike conventional semiconductors or half-metallic ferromagnets, they have a finite band gap for carriers with one spin direction and a zero or nearly zero gap for carriers with opposite directions. They are suitable materials for spintronic devices. The main advantages of SGSs are (1) a high Curie temperature, (2) minimum energy required for the transition of charge carriers from the valence band to the conduction band, (3) a lower carrier concentration in comparison with HMFs and, consequently, the possibility of a simpler control of such carriers. The first experimental observation of the SGS state occurred in 2013 in the inverse Heusler alloy $Mn_2CoAl$. The unique features of the band structure of SGS materials are manifested in their kinetic properties: a relatively large value and a weak temperature dependence of the electrical resistivity, a relatively low carrier concentration, a very small Seebeck coefficient, linear low-temperature magnetoresistivity, etc.

Heusler compounds belonging to the class of topological systems exhibit exotic properties, including those of topological insulators, Weyl semimetals of various types, etc. In addition, magnetic skyrmions can arise in them. In the future, Heusler alloys can become objects on which model calculations and experiments can be carried out to understand the relationship between topology, crystal structure, and various magnetic and electronic characteristics. The possibility of easily tuning the band structure of topological Heusler compounds and, consequently, changing the carrier density and Hall conductivity, allows us to hope for practical applications, e.g., for the implementation of the quantum anomalous Hall effect at room temperature.

It should be emphasized that, in real alloys, it is difficult to strictly distinguish between the HMF, SGS, and/or TSM states. In addition, depending on the alloy composition and external parameters, transitions between these states and ordinary semiconducting and magnetic phases are possible. All this opens up further possibilities for fine control over physical properties.

**Table 1.** Characteristics of complete systems of Heusler alloys Mn$_2$YAl, Fe$_2$YAl, Co$_2$TSi, Co$_2$FeZ (Y is a transition metal, and Z is a s- or p-element): magnetic and electronic states, residual resistivity $\rho_0$, saturation magnetization $M_s$, and spin polarization $P$

| Composition | State | $\rho_0$, μΩ×cm (exp.) | $M_s$, emu/g (exp.) | $P$, % |
|---|---|---|---|---|
| Mn$_2$YAl | | | | |
| Mn$_2$TiAl | — | 305 [62] | 0.2 [62] | 75 [150] calc. |
| Mn$_2$VAl | FM, [62] exp. | 122 [62] | 59 [62] | — |
| Mn$_2$CrAl | — | 250 [62] | — | 92 [150] calc. |
| Mn$_3$Al | Compensated ferrimagnet [62, 144] | 251 [62] | — | 55 [150] calc. |
| Mn$_2$FeAl | Frustrated AFM [155] exp. | 242 [62] | — | 86 [141] calc. |
| Mn$_2$CoAl | SGS [11] exp. | 444 [11] | 18 [62] | 100 [141] calc. |
| Mn$_2$NiAl | FM | 84 [62] | 1.2 [62] | 52 [150] calc. |
| Fe$_2$YAl | | | | |
| Fe$_2$TiAl | FM | 12 [138] | 3.5 [138] | 88 [150] calc. |
| Fe$_2$VAl | Semiconductor [23] exp. | 2020 [138] | — | 0 [150] calc. |
| Fe$_2$CrAl | FM | 443 [138] | 17 [138] | 95 [150] calc. |
| Fe$_2$MnAl | HMF [153] calc. | 263 [138] | 31 [138] | — |
| Fe$_3$Al | FM [154] calc. | 53 [138] | 164 [138] | 54 [150] calc. |
| Co$_2$YSi | | | | |
| Co$_2$TiSi | HMF [71] calc. | 155 [139] | 49 [139] | 24 [71] calc. |
| Co$_2$VSi | HMF [71] calc. | 294 [139] | 6 [139] | 35 [71] calc. |
| Co$_2$CrSi | HMF [71] calc. | 318 [139] | 1 [139] | 80 [71] calc. |
| Co$_2$MnSi | HMF [35] exp. | 16 [139] | 114 [139] | 93 [35] exp. |
| Co$_2$FeSi | HMF [143] calc. | 27 [139] | 97 [139] | 57 [143] calc. |
| Co$_3$Si | FM | 69 [139] | 68 [139] | 66 [142] calc. |
| Co$_2$NiSi | FM | 131 [139] | 46 [139] | 51 [142] calc. |
| Co$_2$FeZ | | | | |
| Co$_2$FeAl | HMF [151] calc. | 50 [138] | 150 [138] | 75 [150] calc. 100 [151] calc. |
| Co$_2$FeGa | FM | 9 [138] | 146 [138] | — |
| Co$_2$FeGe | HMF [154] calc. | 14 [138] | 164 [138] | — |
| Co$_2$FeIn | FM | 2 [138] | 224 [138] | — |
| Co$_2$FeSn | FM | 20 [138] | 97 [138] | — |
| Co$_2$FeSb | FM | 9 [138] | 99 [138] | — |

**Acknowledgments**

We are grateful to our colleagues and co-authors M.I. Katsnelson, Yu.N. Skryabin, N.G. Bebenin, E.I. Shreder, A.V. Lukoyanov, Yu.A. Perevozchikova, V.G. Pushin, and E.B. Marchenkova for valuable discussions, as well as A.N. Domozhirova, A.A. Semiannikova, P.S. Korenistov, and S.M. Emel'yanova for help in completing this review.

**Funding**

This work was carried out as part of the state assignment of the Ministry of Education and Science of the Russian Federation (topic "Spin," no. AAAA-A18-118020290104-2, and "Quantum," no. AAAA-A18-118020190095-4) and was supported by the Russian Foundation for Basic Research (project no. 20-12-50004) and the Government of the Russian Federation (resolution no. 211, contract no. 02.A03.21.0006).